\documentclass[12pt,preprint]{aastex}
\usepackage{natbib}
\usepackage{url}
\usepackage{amsmath}
\usepackage{appendix}
\usepackage{hyperref}
\begin{document}

\title{Observing the Earliest Stages of Star Formation in Galaxies:
$8\mu$m Cores in Three Edge-on Disks}

\author{Bruce G. Elmegreen\altaffilmark{1}, Debra Meloy Elmegreen\altaffilmark{2}}


\altaffiltext{1}{IBM Research Division, T.J. Watson Research Center, 1101 Kitchawan
Road, Yorktown Heights, NY 10598; bge@us.ibm.com}

\altaffiltext{2}{Department of Physics \& Astronomy, Vassar College, Poughkeepsie,
NY 12604}

\begin{abstract}
To study the vertical distribution of the earliest stages of star formation in
galaxies, three edge-on spirals, NGC 891, NGC 3628, and IC 5052 observed by the
{\it Spitzer Space Telescope} InfraRed Array Camera (IRAC) were examined for
compact $8\;\mu$m cores using an unsharp mask technique; 173, 267, and 60 cores
were distinguished, respectively.  Color-color distributions suggest a mixture of
PAHs and highly-extincted photospheric emission from young stars. The average
V-band extinction is $\sim20$ mag, equally divided between foreground and core.
IRAC magnitudes for the clumps are converted to stellar masses assuming an age of
1 Myr, which is about equal to the ratio of the total core mass to the star
formation rate in each galaxy. The extinction and stellar mass suggest an
intrinsic core diameter of $\sim18$ pc for 5\% star formation efficiency.  The
half-thickness of the disk of $8\;\mu$m cores is 105 pc for NGC 891 and 74 pc for
IC 5052, varying with radius by a factor of $\sim2$. For NGC 3628, which is
interacting, the half-thickness is 438 pc, but even with this interaction, the
$8\;\mu$m disk is remarkably flat, suggesting vertical stability. Small scale
structures like shingles or spirals are seen in the core positions. Very few of
the $8\;\mu$m cores have optical counterparts.
\end{abstract}

\section{Introduction}
\label{intro}

{\it Spitzer Space Telescope} InfraRed Array Camera (IRAC) images of M100
\citep[][Paper 1]{elmegreen18} and 15 other spiral galaxies at low inclination
\citep[][Paper 2]{elmegreen19} show regularly spaced $8\;\mu$m peaks in kpc-long
filaments that run along the spiral arms and interarm spurs. The summed IRAC
luminosities of these peaks were found to reproduce the galactic star formation
rates if this phase of young stellar evolution lasts between 0.2 and 2 Myr, which is
reasonable for embedded sources close to their time and place of formation. Other
measures of the star formation rate, such as H$\alpha$ and FUV, get about the same
rates but highlight later phases of evolution when the ionized gas and young stars
break out from their cloud cores \citep{kennicutt12,hannon19}. These later phases,
along with the irregularity of H$\alpha$ emission and the obscuration of optical and
FUV light by dust, hide important clues to the star formation process.

The morphology of $8\;\mu$m peaks presents a much clearer picture. This morphology
suggests that most star formation begins by gravitational collapse in dense gas that
is compressed by large scale dynamical processes, such as spiral arm shocks, spurs,
and giant shells driven by young stellar feedback \citep[e.g.,][Papers
1,2]{elmegreen79,cowie81,tomisaka87,kim01,
kim02a,kim02b,kim06,kim07,dobbs08,renaud13,renaud14,elmegreen14}.

Here we investigate three edge-on galaxies that have IRAC data similar to that in
Papers 1 and 2. We find compact $8\;\mu$m sources using an unsharp mask technique
and study their colors, luminosities and spatial distributions.  Edge-on galaxies
complement our previous study by showing the vertical distributions of the $8\;\mu$m
cores and the thicknesses of the star-forming layers. Also at $8\;\mu$m, the
extinction is so low that we can observe most of the star-forming regions throughout
a galaxy even though the line of sight spans many kiloparsecs.

\section{Observations}
\label{method}

To find edge-on galaxies for this survey, we searched the {\it Spitzer} archives for
IRAC channel 4 ($8\;\mu$m) observations using galaxy samples designed for studies of
thick disks \citep{comeron11,comeron12}. Three good cases were found, NGC 891, NGC
3628, and IC 5052, as listed in Table 1. They are all at a distance of $\sim10$ Mpc
or less, which gives good spatial resolution, and they have $8\;\mu$m sources all
over their disks. Eleven other galaxies in this sample have $8\;\mu$m images but are
either too distant or have too few detectible sources: NGC 1032 and NGC 4441 are too
distant\footnote{https://ned.ipac.caltech.edu/} at 37.1 Mpc and 38.9 Mpc,
respectively. IC610 (15 Mpc), NGC 4081 (21.3 Mpc), and NGC 4111 (11.4 Mpc) primarily
show their nuclear regions at $8\;\mu$m, NGC 4565 (16.8 Mpc) has mostly a nuclear
ring at $8\;\mu$m, and NGC 1495 (15.5 Mpc) and NGC 3592 (16.8 Mpc) do not have much
emission at all at 8 micron. NGC 5981 (26.4 Mpc) has only 3 obvious sources and NGC
3501 (14.5 Mpc) has about 5 sources. NGC 4330 at 20.6 Mpc may be the most
interesting of these additional cases, because it has some dozen detectable
$8\;\mu$m sources in an unsharp mask image, but it is still inferior to the chosen
sample where the number of sources ranges between $\sim50$ and $\sim200$ and the
spatial resolution is 2 or more times better.

\begin{figure}
\includegraphics[width=3.in]{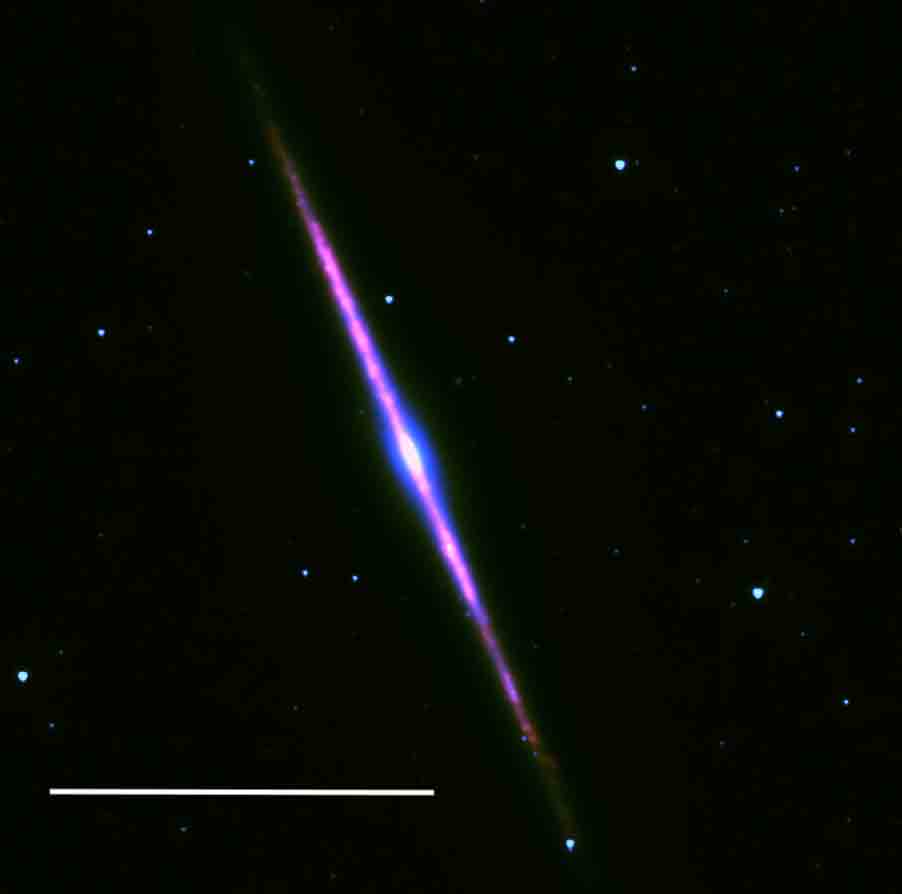}\\
\includegraphics[width=3.in]{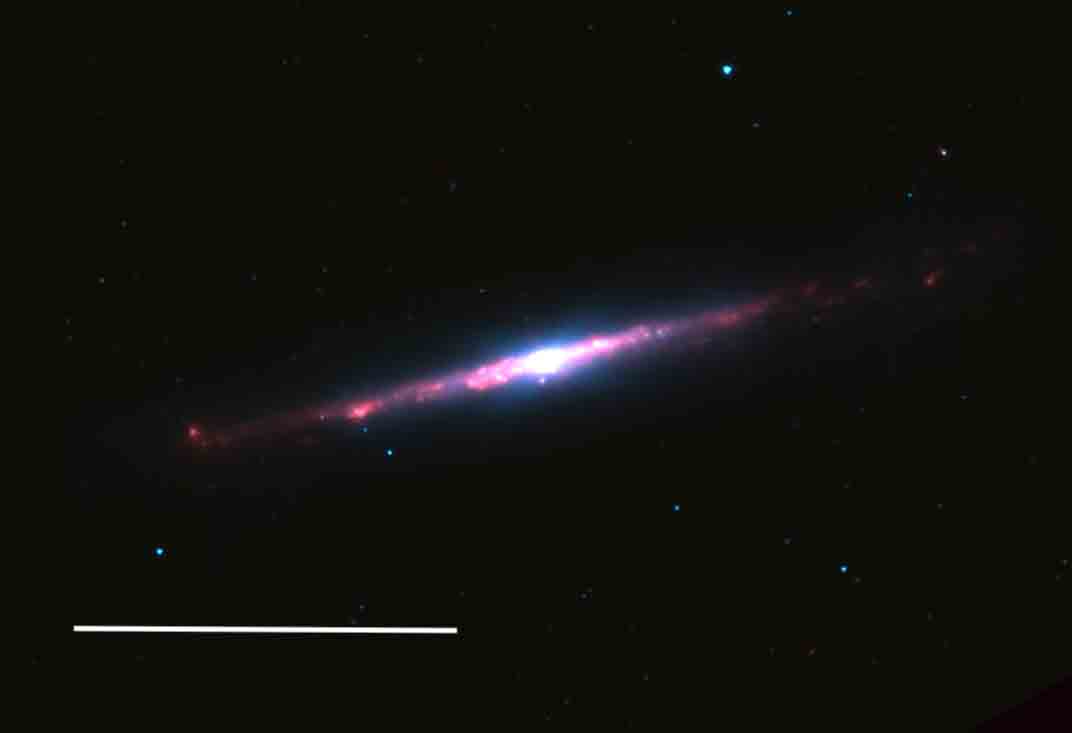}\\
\includegraphics[width=3.in]{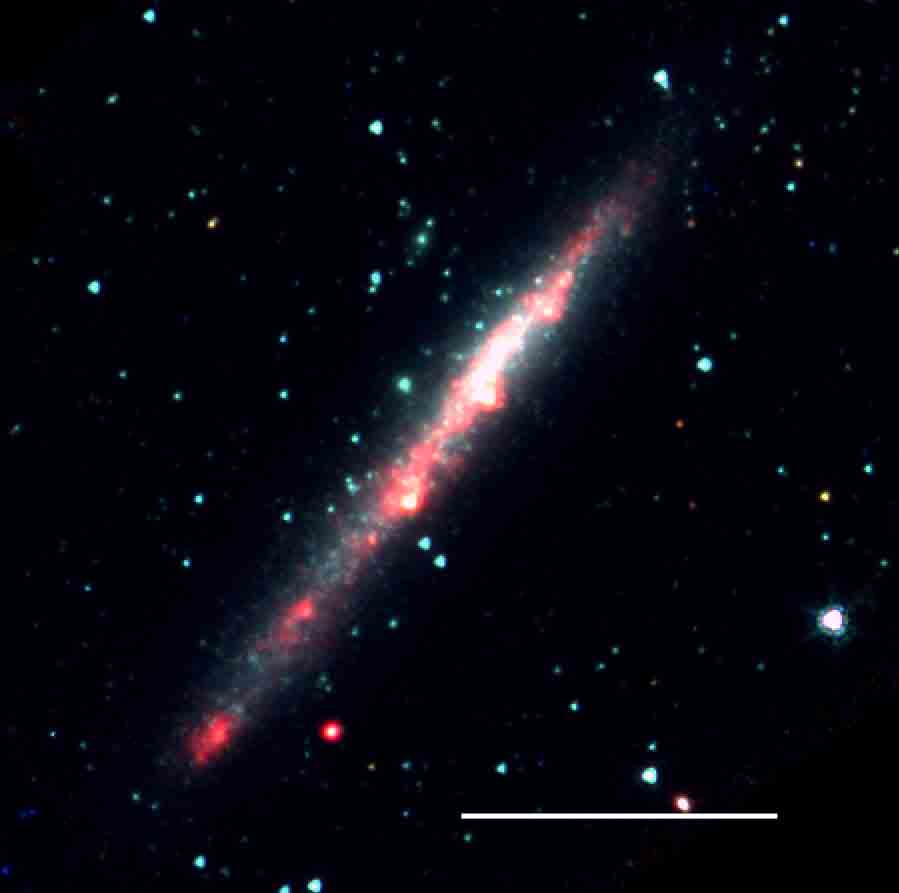}\\
\caption{Color IRAC images of the galaxies in this study from composites
of $3.6\;\mu$m (in blue), $4.5\;\mu$m (in green), and $8\;\mu$m (in red) images. Top: NGC 891
with a scale of 5 arcmin indicated by the line;
middle: NGC 3628 with 5 arcmin indicated;
bottom: IC 5052 with 2 arcmin indicated.
North is up. (Figure degraded for arXiv storage.)}
\label{mycolor_all}
\end{figure}

For the three chosen galaxies, the post-Basic Calibrated Data (BCD) processed images
in all four IRAC bands were retrieved as FITS files and aligned through the Image
Reduction Analysis Facility (IRAF) task {\it wregister}. Figure \ref{mycolor_all}
shows them with color IRAC images made using the software package DS9 (north is up,
west is to the right). Each is a composite of the archived $3.6\;\mu$m (in blue),
$4.5\;\mu$m (in green), and $8\;\mu$m (in red) images.  The pixel size in the IRAC
mosaics used here is $0\farcs75$, and the angular resolution at $8\;\mu$m
is $2\farcs4$ FWHM \citep{chambers09}. Other angular resolutions and
processing data are in Table \ref{photometry}.

\begin{figure}
\epsscale{1.}
\includegraphics[width=6.in]{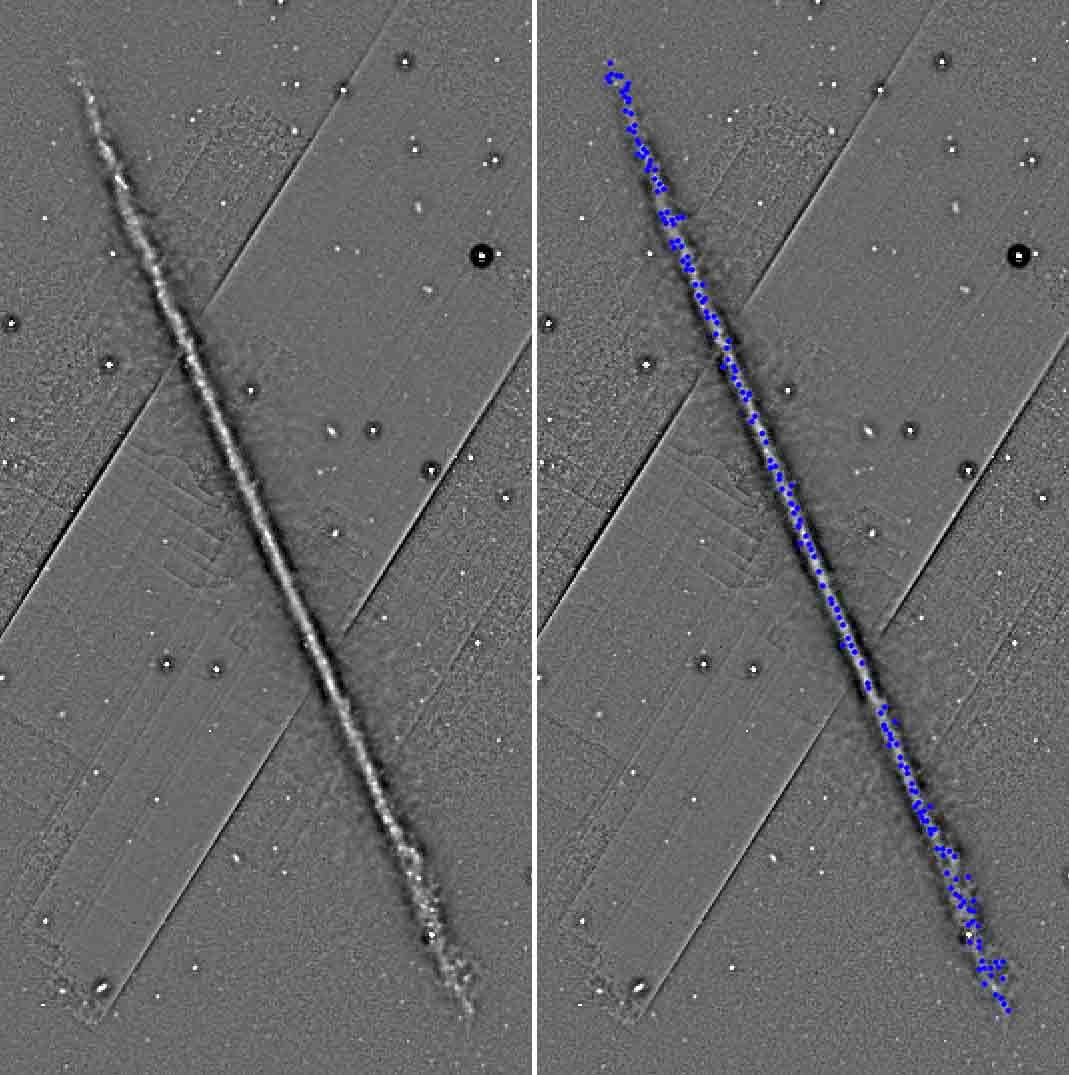}
\caption{Unsharp mask image of NGC 891 (left) made by dividing the IRAC $8\;\mu$m image
by a Gaussian blurred version of itself, using a Gaussian $\sigma=3$ pixels,
or $2\farcs25$.
173 $8\;\mu$m cores identified on the image and listed in Table \ref{listofclumps891} are shown on the
right as blue dots. (Figure degraded for arXiv storage.)}
\label{n891_usm_dual_v3}
\end{figure}

\begin{figure}
\epsscale{1.}
\includegraphics[width=6.in]{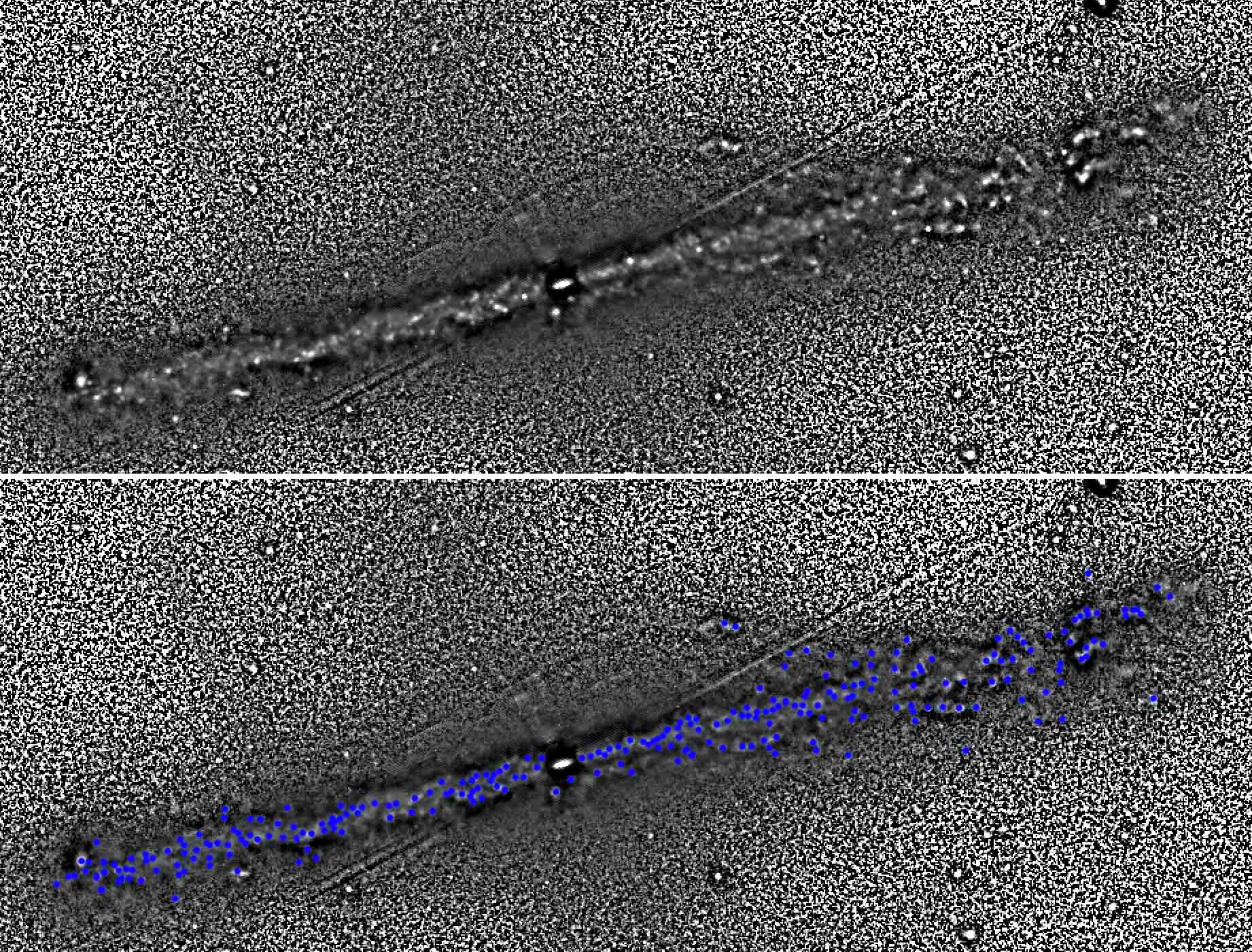}
\caption{Unsharp mask of the $8\;\mu$m image of NGC 3628 (top) with 267
cores from Table \ref{listofclumps3628} shown as blue dots (bottom). North is up.
Negative positions in Figure \ref{efremov_edgeon_xy2}
are on the right here. (Figure degraded for arXiv storage.)
}
\label{n3628_usm_dual_v2}
\end{figure}

\begin{figure}
\epsscale{1.}
\includegraphics[width=6.in]{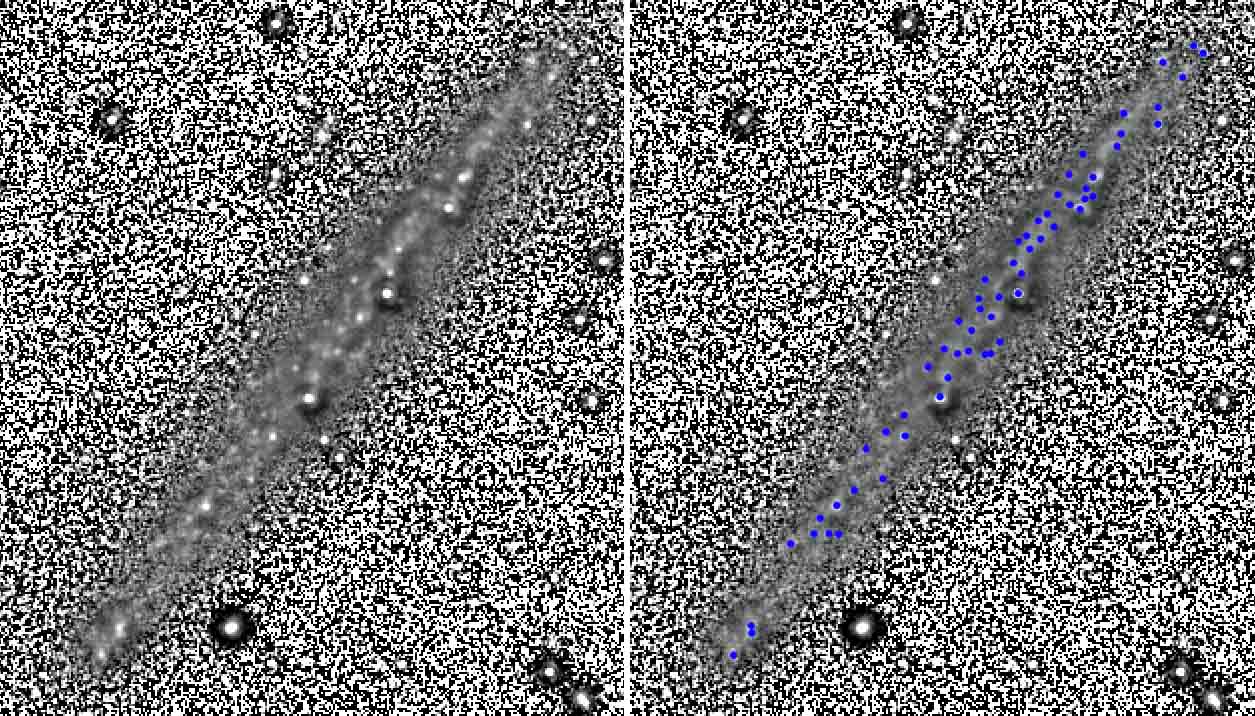}
\caption{Unsharp mask of the $8\;\mu$m image of IC 5052 (left) with 60
cores from Table  \ref{listofclumps5052} shown as blue dots (right). (Figure degraded for arXiv storage.)}
\label{ic5052_usm_dual_v3}
\end{figure}

Unsharp mask images were made by dividing the $8\;\mu$m image by a blurred version
of the same $8\;\mu$m image using the IRAF task ${\it gauss}$ with a 3 pixel sigma.
Figures \ref{n891_usm_dual_v3}-\ref{ic5052_usm_dual_v3} show the results. On the
left are the unsharp mask images in grayscale and on the right are the same images
with the identified cores superimposed as blue dots. Most of the $8\;\mu$m cores in
the unsharp mask images can also be seen as diffuse peaks in the IRAC color images.
Division by the blurred $8\;\mu$m image highlights the densest emission regions and
normalizes all the local backgrounds to show them in a single image.  The selection
effect for 6 pixels diameter in the unsharp mask procedure (i.e., twice the Gaussian
sigma of the mask) implies that the luminosities and other derived quantities are
limited to the core regions of what could be larger objects. Six pixels is still
large compared to giant molecular clouds, however.  The spatial scale of 6 pixels is
199 pc, 225 pc, and 120 pc for NGC 891, NGC 3628 and IC 5052, respectively, whereas
giant molecular clouds are typically 20 to 50 pc in size \citep{sanders85}. Thus,
the masks presumably capture most of the luminosity associated with localized star
formation.

\section{Properties of $8\;\mu$m Cores}
\label{properties}

Magnitudes of the $8\;\mu$m peaks on the original FITS images were determined in the
four IRAC bands using the IRAF task {\it phot} with a measurement aperture of 2
pixels radius ($3\farcs0$ diameter), and background subtraction from an
annulus between 3 and 4 pixels radius. The zeropoints and aperture corrections for
conversion of counts to Vega-system magnitudes in each filter were taken from the
IRAC Instrument
handbook\footnote{http://irsa.ipac.caltech.edu/data/SPITZER/docs/irac/irac
instrumenthandbook/}  and are given in Table \ref{photometry}.

The right ascension, declination, magnitude and statistical error in magnitude for
each filter are listed in Tables \ref{listofclumps891} to \ref{listofclumps5052},
along with the core colors that equal the magnitude differences. Blank entries
represent core measurements that have statistical errors larger than 3 magnitudes or
which could not be detected. We also removed $8\;\mu$m sources with
$[5.8]-[8.0]<0.6$, which are probably stars \citep {allen04}.  Note that cores with
a $4.5\;\mu$m detection and no $3.6\;\mu$m detection have large error limits in the
$4.5\;\mu$m band, making the $3.6\;\mu$m band, which is generally fainter because of
extinction (see below), undetected. The $8\;\mu$m band detects the most sources
because it has the least extinction and has excess emission from PAHs; this is why we
searched for sources on an unsharp mask image at $8\;\mu$m. All of the Table 1
entries have a blue dot in Figures \ref{n891_usm_dual_v3}-\ref{ic5052_usm_dual_v3}.

\begin{figure}
\epsscale{1.}
\includegraphics[width=6.in]{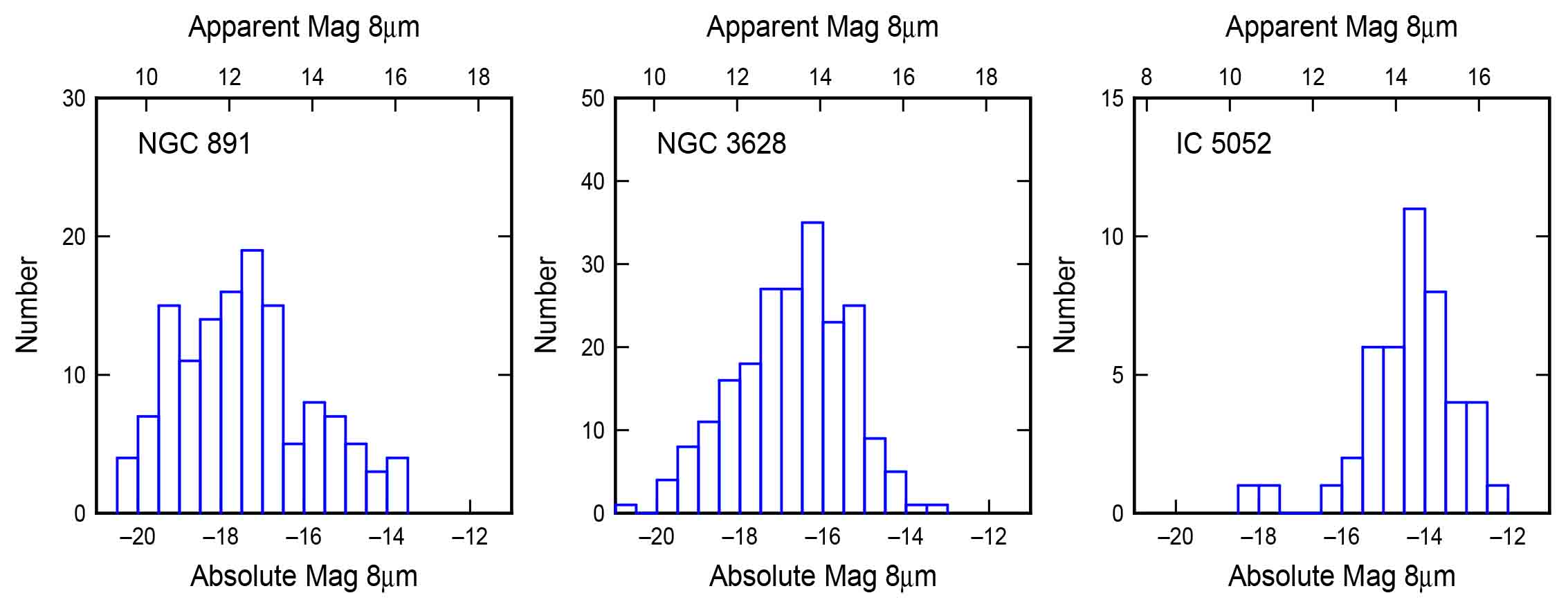}
\caption{Distribution function of absolute and apparent (top axis) magnitudes
at $8\;\mu$m of the cores found in unsharp mask images, as measured on the
original images. The Vega magnitude scale is used. The two brightest cores in
IC 5052, which stand out to the left in the distribution, are also highlighted
with red circles in Fig. \ref{efremov_edgeon_colors} and correspond to bright blue
optical regions in the color image of Fig. \ref{Hubble_I5052_triplet} (Figure degraded for arXiv storage.).}
\label{efremov_edgeon_mags}
\end{figure}

The distributions of $8\;\mu$m absolute and apparent magnitudes for sources that are
detected in all four IRAC bands are shown in Figure \ref{efremov_edgeon_mags}. The
cores typically become undetected below approximately $-17$ mag, $-16$ mag and $-14$
mag for NGC 891, NGC 3628 and IC 5052. Larger-scale apertures for the photometry
would include slightly more flux (e.g., Paper 2).

\begin{figure}
\epsscale{1.}
\includegraphics[width=6.in]{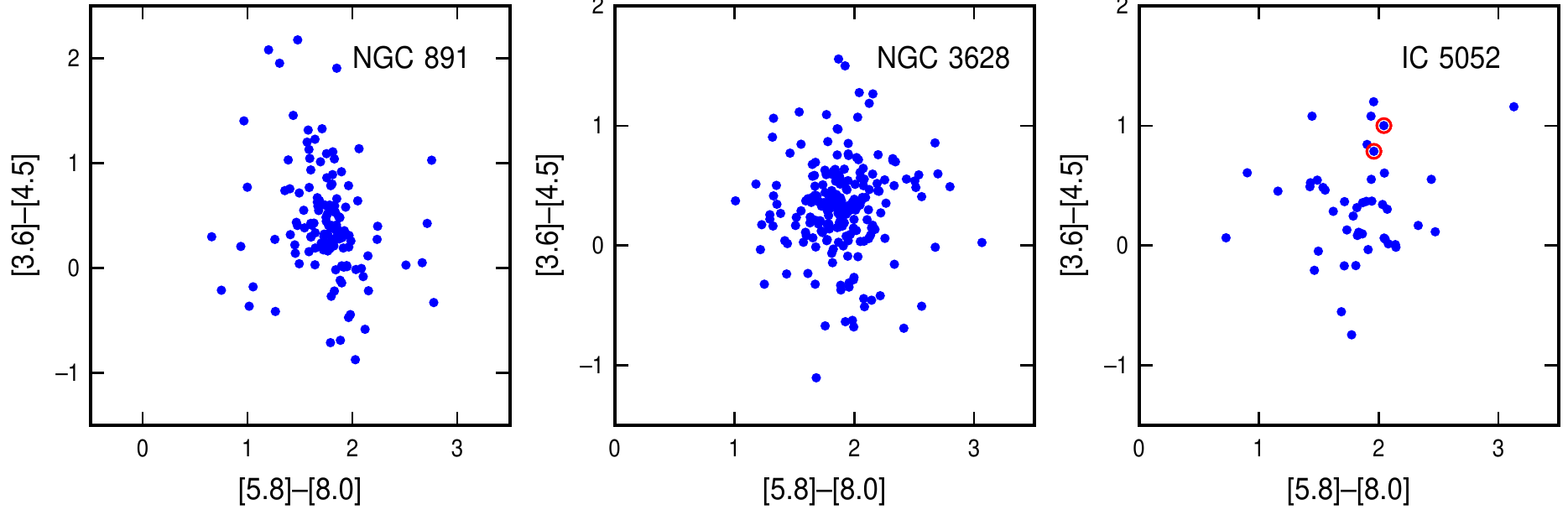}
\caption{Distribution of IRAC colors for $8\;\mu$m cores on the Vega scale. The excess
of the $[3.6]-[4.5]$ color above 0 is from extinction and hot dust.
The $[5.8]-[8.0]$ color is typical of whole galaxy disks and probably dominated by PAH emission.
The red circles in the plot for IC 5052 are the two bright optical star forming
regions discussed in Section \ref{color5052}.
}
\label{efremov_edgeon_colors}
\end{figure}

Region colors, shown in Figure \ref{efremov_edgeon_colors}, are not affected much by
the aperture size (the points with red circles for IC 5052 will be discussed in
Section \ref{color5052}).  The average values for the Vega-system colors and their
dispersions $\sigma$ are in Table \ref{colors}. Individual color uncertainties from
measurement errors average 1.5 mag, but these are random so their contribution to
the uncertainties in the average colors is approximately their ratio to the square
root of the number of cores, or $\pm0.008$.

The average for $[3.6]-[4.5]$ is $0.35\pm0.02$.  In Paper 2, we found an average of
$[3.6]-[4.5]=0.21\pm0.002$ mag for 8300 cores in 15 face-on galaxies. This lower
value for face-on galaxies is presumably for the star forming cores alone, and the
extra $\sim0.14$ mag color excess for the present sample is from the long columns of
dust on the lines of sight to the cores.

The colors suggest that emission from extincted photospheres makes $[3.6]-[4.5]$
positive but low, and emission from PAHs makes $[5.8]-[8.0]$ high \citep[Papers
1,2;][]{elmegreen06}. Bare photospheric emission has $[3.6]-[4.5]\sim0$ in the Vega
system \citep{allen04}, so our average value of $[3.6]-[4.5]=0.35\pm0.02$ implies
contributions from extinction, PAH emission, and hot dust emission. According to
\cite{megeath04}, a visual extinction of 30 mag produces a $[3.6]-[4.5]$ color
excess between 0.4 and 0.45 depending on the extinction law.   Then $A_{\rm
V}=30/0.425=71([3.6]-[4.5])$. \cite{chapman09} measured $A_{\rm 3.6}/A_{\rm K_{\rm
s}}=0.64\pm0.03$ and $A_{\rm 4.5}/A_{\rm K_{\rm s}}=0.53\pm0.03$ for local molecular
clouds at $A_{\rm K_{\rm s}}>2$, which is appropriate for our sources, and
\cite{bessell88} found $A_{\rm K}/A_{\rm V}=0.109$. Together these give $A_{\rm
V}=([3.6]-[4.5])/(0.109[0.64-0.53])=83([3.6]-[4.5])$. \cite{flaherty07} got about
the same for local clouds. We use $A_{\rm V}=80([3.6]-[4.5])$ here, and then
$[3.6]-[4.5]\sim0.35$ corresponds to a visual extinction of $\sim28$ mag if there is
no emission from PAHs and hot dust.

Hot dust can cause $[3.6]-[4.5]$ to increase above the value from extinction alone.
\cite{mentuch10} measured the ratio of intensities at $4.5\;\mu$m and $3.6\;\mu$m
for individual pixels in 56 local galaxies and plotted this ratio versus the
dust-corrected intensity of H$\alpha$ as a measure of local UV radiation field. The
$4.5\;\mu$m to $3.6\;\mu$m ratio increased with H$\alpha$ approximately as $\Delta
\log(I[4.5]/I[3.6])\sim0.05 \log(I[H\alpha])$ for $I(H\alpha)$ in MJy sr$^{-1}$.
\cite{mentuch10} suggested that this increase is from a combination of dust heating,
which increases the intensity at longer wavelengths, and PAH emission in the
$3.6\;\mu$m band. We have no H$\alpha$ measurements of our $8\;\mu$m cores, but can
estimate the intensity from the embedded cluster masses derived below, which are on
the order of $\sim10^4\;M_\odot$. For a population of stars younger than 1 Myr with
a fully sampled Salpeter IMF, as assumed for the mass derivation, the Lyman
continuum emission rate would be $10^{46.4611}$ s$^{-1}$ per solar mass of stars
\citep{bruzual03}, and, using the conversion in \cite{kennicutt98}, the H$\alpha$
luminosity would be $4.0\times10^{34}$ erg s$^{-1}$ per solar mass. We convert this
to intensity by dividing by the product of a presumed H$\alpha$ linewidth of 30 km
s$^{-1}$ converted to a frequency interval, the solid angle of the source and $4\pi$
times the square of the distance. The result is an estimated H$\alpha$ intensity, in
MJy sr$^{-1}$ of $I(H\alpha)=230(M_{\rm stars}/R^2)$ for stellar mass $M_{\rm
stars}$ in $M_\odot$ and source radius $R$ in pc. Then the \cite{mentuch10}
observation implies $[3.6]-[4.5]\sim0.125\log(230M/R^2)$ from hot dust and PAH
emission if all the radiation from that stellar mass goes to heat the dust. A
correction to this estimate is needed for young stellar clusters with masses less
than $10^3M_\odot$ because they are not likely to have a completely sampled IMF.
Using tables in \cite{vacca96} for a Salpeter IMF, we estimate that the H$\alpha$
luminosity per solar mass starts to drop quickly below $10^3\;M_\odot$.

For $M_{\rm stars}\sim10^4\;M_\odot$ and $R\sim 100$ pc from the physical resolution
of our survey, the above result implies $[3.6]-[4.5]$ should increase by $0.3$
because of hot dust. This is the full color excess we observe and suggests there is
no extinction, which is unreasonable. More likely, only a fraction of the stellar
luminosity that makes the $8\;\mu$m emission is heating the dust to a high
temperature, and the rest is distributed to lower-density surrounding regions. If we
assume $\sim5$\% of the luminosity is responsible for the most heating, then the
excess $[3.6]-[4.5]$ would be $\sim0.1$. Obviously this correction for hot dust is
highly uncertain. We also assume the average color excess for cores in the three
edge-on galaxies, $[3.6]-[4.5]=0.35\pm0.02$, comprises a component from the core
itself equal to $[3.6]-[4.5]=0.21\pm0.002$, as determined for face-on galaxies in
Paper 2, plus a foreground color excess from extinction equal to the residual,
$[3.6]-[4.5]=0.14\pm0.02$.  Then, with a component of $[3.6]-[4.5]\sim0.1$ mag from
local heating, the sources are left with $[3.6]-[4.5]\sim0.11$ mag from extinction
alone. That corresponds to $A_{\rm V}\sim8.8$ mag using the above calibration.

The gas mass surface density corresponding to this extinction comes from the
conversion of $A_{\rm V}$ into HI column density. The local conversion factor
between color excess and HI column density is $E(B-V)=N(HI)/(5.8\times10^{21}\; {\rm
cm}^{-2})$ \citep{bohlin78}, and the ratio of total-to-selective extinction for
diffuse gas is $A_{\rm V}/E(B-V)=3.1$, giving $N(HI)=1.87\times10^{21}A_{\rm V}$.
\cite{wolk08} got the about same,  $N(HI)=2.0\times10^{21}A_{\rm V}$, from X-ray and
IR observations of a massive star-forming region. \cite{winston10} however got
$N(HI)=0.89\pm0.12\times10^{22}A_{\rm K}=0.97\pm0.13 \times10^{21}A_{\rm V}$
cm$^{-2}$ from X-ray and IR observations. Similarly, \cite{chapman09} found $R=5.5$
at $A_{\rm K}>1$ mag, which would lower the \cite{bohlin78} conversion to
$N(HI)=1.1\times10^{21}A_{\rm V}$. Here we take $N(HI)=1.0\times10^{21}A_{\rm V}$
for the highly extincted regions observed as $8\;\mu$m emission cores. This HI
column density converts to a mass surface density of $\Sigma_{\rm gas}=10.8A_{\rm
V}\;M_\odot$ pc$^{-2}$ assuming a mean atomic weight of 1.36. Thus, $A_{\rm
V}\sim8.8$ mag from the corrected $[3.6]-[4.5]$ color excess corresponds to a mass
surface density close to the source of $95\;M_\odot$ pc$^{-2}$. This is comparable
to the surface densities of giant molecular clouds \citep{heyer15}.

Considering now the foreground extinction, the difference, mentioned above, between
the average $[3.6]-[4.5]=0.35\pm0.02$ found here for $8\;\mu$m cores in three
edge-on galaxies and the average $[3.6]-[4.5]=0.21\pm0.002$ for 15 face-on galaxies
in our previous study is $[3.6]-[4.5]=0.14\pm0.02$. This is presumably from diffuse
dust on the line of sight, so we convert it to extinction without the correction for
hot dust, using $A_{\rm V}=80([3.6]-[4.5])$ from above. The result is $A_{\rm
V}=11.2$ mag, which corresponds to a foreground column density
$N(HI)=2.2\times10^{22}$ cm$^{-2}$ for the calibration $N(HI)=2.0\times10^{21}A_{\rm
V}$ (the value for normal regions). Dividing this foreground column density by a
typical pathlength of $\sim5$ kpc through the edge-on disk, the average space
density is $1.5$ H cm$^{-3}$, which is a reasonable value for the midplane of a
spiral galaxy \citep{bohlin78}. Alternatively, the mass surface density of
foreground gas is $\sim240\;M_\odot$ pc$^{-2}$ from the conversion $\Sigma_{\rm
gas}=21.6A_{\rm V}\;M_\odot$ pc$^{-2}$ discussed above, again with a factor of 2
multiplier for normal regions.  Dividing this by 5 kpc gives $0.04\;M_\odot$
pc$^{-3}$, which is about the same as the average gas density in the solar
neighborhood, $0.043\pm0.004\;M_\odot$ pc$^{-3}$  \citep{mckee15}.

Some $8\;\mu$m cores have much higher $[3.6]-[4.5]$ colors than average, greater
than 1 in NGC 891 and NGC 3628 (Fig. \ref{efremov_edgeon_colors}). The corresponding
extinction could be several times higher than the average for individual cores. In
comparison, there were hardly any cores with $[3.6]-[4.5]>1$ in the face-on
galaxies. The extremely red regions in the edge-on cases are probably blends from
several star-forming regions on the line of sight.

Blends are expected in edge-on galaxies for lines of sight that are close to the
center or parallel to spiral arms.  Paper 2 found 8300 $8\;\mu$m cores brighter than
a completeness limit in 15 face-on galaxies, for an average of $\sim500$ detected
$8\;\mu$m cores per galaxy. Here we find 500 distinct cores in 3 edge-on galaxies,
for an average of 170 per galaxy. That difference suggests we are missing two-thirds
of the cores because of excessive faintness or blending. The resolution of our
survey from the unsharp mask is $R_{\rm resol}\sim100$ pc, so each object projects
an area $\pi R_{\rm resol}^2$ in a disk of area $4R_{\rm disk}H$ for half-thickness
$H$ and projected disk radius $R_{\rm disk}$. Taking $R_{\rm disk}\sim10$ kpc and
$H\sim100$ pc, the ratio of core area to disk area is 0.8\%. Considering Poisson
probabilities with 500 potential cores, the fraction of the projected area of the
galaxy without a core is only $\exp(-0.008*500)=2$\%. Thus most cores overlap with
some other core, especially in the denser regions.

\begin{figure}
\epsscale{1.}
\includegraphics[width=6.in]{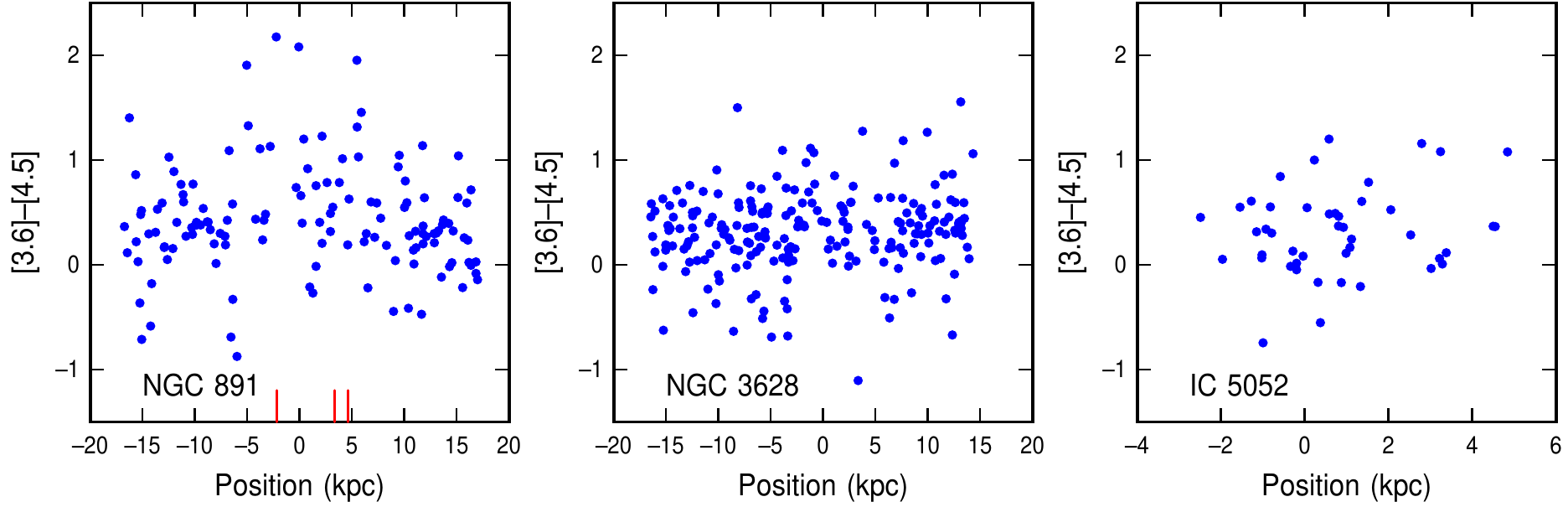}
\caption{The $[3.6]-[4.5]$ color for each core, taken as a measure of extinction for young stellar
photospheric emission, is plotted versus the position along the midplane of the galaxy, with
zero position corresponding to the center observed in IRAC channel 1 ($3.6\;\mu$m).
$[3.6]-[4.5]$ is
fairly uniform for cores across the disks, although there is a lot of variation in
NGC 891 with a slight excess near the center, most likely corresponding to blending on the
line of sight. Negative position is to the west, which for NGC 891 is mostly to the
south and for IC 5052 is mostly to the north. The red lines at the bottom of the panel
for NGC 891 are at the positions of strong HI and molecular features that could be
spiral arms viewed tangentially.}
\label{efremov_edgeon_color-position}
\end{figure}

Figure \ref{efremov_edgeon_color-position} shows the $[3.6]-[4.5]$ colors versus
position along the galaxy in kpc, with position zero being the galactic center in
the $3.6\;\mu$m IRAC image and negative position to the west. The colors are
slightly redder near the center for NGC 891, but not for the other galaxies. Red
lines at the bottom of the panel for NGC 891 show the positions of strong features
in CO(3-2) that could be inner spiral arms \citep{dumke01}.  There is no obvious
additional reddening of the cores in these spiral regions.

The average optical extinction derived from IRAC colors, $\sim8.8$ mag, is about the
same as for giant molecular clouds in the Milky Way \citep{heyer15}, suggesting we
are seeing the bright regions associated with giant molecular clouds where clusters
and OB associations form. Such cores would be much smaller than the pixel size of
$0\farcs75$. We show below that the average young stellar mass is within a
factor of a few either way times $10^4\;M_\odot$. If the total gas mass is
$20\times$ the stellar mass for a local star formation efficiency of 5\%, and the
gas mass surface density in front of the stars is $\sim95\;M_\odot$ pc$^{-2}$, as
suggested above from the source extinction, and if there is an equal gas surface
density associated with the cluster behind it, then the average diameter of a core
is $\sim18$ pc. This is similar to that of molecular and infrared-dark clouds in the
Milky Way \citep{simon06} and a factor of $\sim3$ larger than young clusters in
other galaxies \citep{ryon17}.

The $[5.8]-[8.0]$ color is not significantly affected by extinction \citep{allen04},
so the mean value of $[5.8]-[8.0]=1.83\pm0.02$ for our sources requires an excess
emission at $8\;\mu$m. We discussed in Paper 2 how this excess, which resembles that
in integrated galaxy disks \citep{stern05,winston07,gutermuth09,stutz13}, is likely
to be from PAH emission on carbonaceous grains, which produce  emissions in both
bands with an excess at $6\mu$m to $10\mu$m \citep{li01}.

\section{Positional Distribution of the $8\;\mu$m cores}

\subsection{Disk Thickness}

The distribution function of core distances from the midplane is shown for each
galaxy in Figure \ref{efremov_edgeon_xy2}. The midplane was determined by fitting
the $8\;\mu$m core positions to a line. The rms dispersion of the midplane distance,
considered to be the half-thickness of the disk distribution of $8\;\mu$m cores, is
105 pc for NGC 891, 438 pc for NGC 3628, and 74 pc for IC 5052.  For NGC 891 and IC
5052, these are good representations of the disk half-thickness, but NGC 3628 is
warped as a result of a recent interaction \citep[it is part of the Leo triplet, Arp
317,][]{arp66}, and it is much thicker on the western side.

Figure \ref{efremov_edgeon_xy2} shows the distance to the midplane for each core as
a function of projected galactocentric radius. In the bottom right panel, which is
for NGC 3628, the distance to the midplane is plotted versus position along the
midplane, with position 0 being the center of the galaxy as determined by the peak
in the $3.6\;\mu$m emission. Negative position for NGC 3628 is to the northwest,
where there is a large warp and thickening on deep HI \citep{wilding93} and optical
images. The red crosses in the figure are average distances for positions in the
intervals centered around them. NGC 891 has a steadily increasing distance to the
midplane that varies as
\begin{equation}
H({\rm pc})=(49.9\pm32.9)+(6.3\pm3.1)\times R({\rm kpc}),
\label{eq:thick891}
\end{equation}
which is from $\sim71$ pc inside 6 kpc (the first three crosses) to 142 pc outside
12 kpc. IC 5052 has a nearly constant, or possibly decreasing, midplane distance,
varying as
\begin{equation}
H({\rm pc})=(123\pm29)-(15.4\pm14.8)\times R({\rm kpc})
\end{equation}
with an average inside 3.5 kpc of 100 pc. The southeast part of NGC 3628, which is
relatively free of the warp, has a midplane distance that is constant to within the
errors,
\begin{equation}
H({\rm pc})=(241\pm219)+(2.68\pm24.6)\times R({\rm kpc}).
\label{eq:thick3628}
\end{equation}
These trends are shown in Figure \ref{efremov_edgeon_xy2}.  A slight increase with
radius is expected in a typical galaxy \citep{narayan02} and is observed in Milky
Way atomic \citep{malhotra95} and molecular \citep{heyer15} gas beyond about half
the galactocentric radius of the Sun.

The trend of disk half-thickness with radius can be a check on galaxy inclination. A
galaxy that is not exactly edge-on will have a larger projected thickness near the
center than near the edge because of the higher extension of the near and far parts.
In Figure \ref{efremov_edgeon_xy2}, NGC 891 has a low projected thickness near the
center and a high inclination, $89\fdg7$.  However, IC 5052 has a high projected
thickness near the center and also a high inclination, $\sim90^\circ$. Thus the
decrease in thickness with radius for IC 5052 is not an inclination effect.

The half-thickness of the $8\;\mu$m core distribution is comparable to the thickness
of the interstellar medium in typical disk galaxies. For NGC 891, \cite{scoville93}
measured a half-thickness of CO emission that ranged from 77 pc in the center to 132
pc at 8.8 kpc (adjusted to our distance). This is 30\%-50\% larger than the
half-thickness of the $8\;\mu$m core distribution given above, which ranges from 50
pc to 105 pc in the same radial range (from the linear fit). \cite{alton00} modelled
$850\mu$m observations of this galaxy and found that the dust follows the molecular
gas in the main part of the disk.

In addition to a thin interstellar disk where stars form, some galaxies have a thick
interstellar disk that may be visible as dust filaments or H$\alpha$ emission
\citep{howk99}. \cite{jo18} observed extended H$\alpha$ and FUV emission from 38
nearby edge-on galaxies and found a thick disk component in around half of them,
including NGC 891 and NGC 3628, and only a thin component in IC 5052. In NGC 891 and
NGC 3628, the H$\alpha$ scale height is $\sim850$ pc and the FUV and NUV scale
heights are about twice that. In IC 5052, the H$\alpha$ scale height is $\sim350$ pc
and the FUV and NUV heights are about the same. They suggested that the great
heights for H$\alpha$ indicate an extended component of gas and dust that is ionized
and also scatters midplane radiation. They found that the relative thickness of the
extraplanar gas, normalized to the size of the disk, scales with the star formation
rate per unit area to a power between 0.3 and 0.5, depending on whether extended PAH
emission is included. NGC 891 also has a thick component of HI with a scale height
of 2.2 kpc ($50^{\prime\prime}$), and radio continuum emission with a height of 1.1
kpc; these are in addition to a thin HI disk with a height less than 300 pc
\citep{oosterloo07} that is presumably connected with our $8\;\mu$m core disk. At
100, 160 and 250 $\mu$m in NGC 891, the scale heights are $0.24\pm0.05$,
$0.43\pm0.06$ and $1.40\pm0.24$ kpc \citep{hughes14}. The extended dust layer could
be formed by accretion \citep{hodges18} or galactic fountains powered by star
formation \citep{bregman13}.

\cite{shinn15} modelled radiative transfer of FUV emission in NGC 891 and NGC 3628
with a two-component disk for dust scattering, and \cite{baes16} modeled NGC 3628
with the same two components. For NGC 891, the thin dust disk was modeled to have a
scale height of 325 pc and the thick dust scale height was 2100 pc. For NGC 3628,
the thin and thick scale heights were modeled to be 225 pc and 2100 pc.
\cite{bianchi08} modeled radiative transfer in NGC 891 using an extended dust disk
and a clumpy dust disk with a scale height of 200 pc. \cite{popescu11} fit radiative
transfer models to the SED in NGC 891 with three dust components, one that follows
the old stars, another that has the same half-thickness as young stars, assumed to
be 90 pc, and a third that is clumpy with the same distribution as the second. This
clumpy dust model is consistent with the distribution of $8\;\mu$m cores.

\subsection{Shingles and Spiral Arms}

A close examination of the distribution of $8\;\mu$m cores shows corrugations and
shingle-like structures, as observed in the Milky Way \citep[see review
in][]{alfaro96} and other galaxies \citep{florido91}. Figures \ref{n891_separated}
to \ref{ic5052_separated} show enlargements of the unsharp mask images so that
individual cores can be seen. Aside from seemingly random positions above and below
the midplane for most of the cores, especially near the ends of the disks where the
scale height is large, there are also streaks of clumpy emission, such as one in the
center of the lower panel in Figure \ref{n891_separated} for NGC 891. NGC 3628 in
Figure \ref{n3628_separated} has a chevron-type structure on the right-hand side of
the lower panel (just to the left of the nuclear disk as indicated by the arrow)
with three symmetric streaks of multiple cores above and below the plane. In the
streaks closest to the center, which are the best defined, the lower one has three
cores with a bright one in the middle, and the upper one has three cores with equal
brightness. From the coordinates of the cores, we determine the projected lengths of
these two streaks to be $12.3$ pixels ($=9\farcs2=460$ pc) and $11.0$
pixels ($=8\farcs2=410$ pc); the projected angles relative to the midplane
are $40\farcs5$ and $21\fdg4$, respectively. Similarly inclined streaks of
multiple $8\;\mu$m clumps are in IC 5052 (Fig. \ref{ic5052_separated}); the top
panel shows at least two of them, sloping upward and to the right.

Some of these clump alignments could be spiral arms that lie in the plane of the
galaxy. Papers 1 and 2 showed that most spiral arms have $8\;\mu$m cores lined up on
narrow dust filaments that can extend for several kpc. For M100, the average core
separation along the filaments is 410 pc. NGC 3628 is the least inclined galaxy
($i=79\fdg3$ in Table 1) and thus the one most likely to show spiral arms. The
chevron-like structures could then be spirals viewed tangentially. For a single
triplet of projected length $L=410$ pc, which is the second one mentioned above, at
an angle to the midplane of $\theta=21\fdg4$, the filament length in the plane is
$L(\cos[\theta]^2+\sin[\theta]^2/cos[i]^2)^{0.5}=892$ pc. The separation between the
cores in the triplet is half of this, $445$ pc, like the separations in M100. Note
that the chevrons or spirals can also be seen in Figure 1b to the left of the bright
central region. The distance between the apex of the inner chevron and the galaxy
center is 1.9 kpc, which would be the galactocentric distance of the spiral arm.

Generally the galaxy disks in Figures \ref{n891_separated} to \ref{ic5052_separated}
show small irregularities around the midplane. Corrugations like this in the Milky
Way have a variety of scales, including vertical oscillations of atomic and
molecular gas that span several kpc in the radial direction and $\sim50$ pc in
height \citep{malhotra95}. Smaller scale vertical motions connected with spiral arms
are observed with H$\alpha$ velocities in other galaxies on a scale of several
hundred pc; they suggest density wave forcing with possible magnetic field effects
\citep{martos98,sanchez-gil15}. Usually the corrugations are only in the gas or
young stellar populations \citep{matthews08b}, as we see here for the $8\;\mu$m
cores. Milky Way ``shingles'' are linearly rising corrugations with an abrupt jump
down to the next one in a sequence \citep{schmidt-kaler93}. Shingles have lengths of
$\sim1$ kpc, thicknesses less than 70 pc and inclinations to the plane of
$\sim10^\circ$ \citep{quiroga77,kolesnik79}. The chevrons in NGC 3826, with
symmetric pairs positioned on each side of the midplane, do not resemble the
vertical structures in the Milky Way or other galaxies, which supports their
identification with spiral arms. These spirals are not visible in the optical image
(Sect. \ref{optical3628}).

There are also loops with several $8\;\mu$m cores suggestive of pressurized bubbles
with gravitational condensations and star formation. Figure \ref{n891_separated}
shows one in the middle of the top panel for NGC 891. Figure \ref{n3628_separated}
shows several seemingly coherent structures above and below the disk of NGC 3628 in
the top panel, which is the western part where the disk thickens most.

\section{Comparison of optical and $8\;\mu$m images}

\subsection{NGC 891}

The optical structure of NGC 891, divided into two sections, is compared to the
$8\;\mu$m unsharp mask image and to the positions of the catalogued $8\;\mu$m cores
in Figure \ref {N891APOD_ch4divg3_triplet_thin_double_v2}. The north-east end of the
galaxy is to the left in the bottom panel. The alignment was made using stars
visible outside the disk in both images, not shown in the figure. The $8\;\mu$m
cores are indicated by circles through which the optical image can be seen. Most of
the circles contain no optical emission, presumably because the dust extinction on
the line of sight to the cores is too high. This is consistent with our findings for
M100 (Paper 1), a more face-on galaxy, suggesting that even the local extinction at
the source is enough to obscure young stars. Some point sources are evident in both
images of NGC 891, however, such as the central one of the shingle-like structure in
the northeast, which is on the left in the lower panel of Figure \ref
{N891APOD_ch4divg3_triplet_thin_double_v2}; this object looks very red in the
optical image. Also, a few faint spots in a high latitude loop above the
southwestern end can be seen in the optical image, inside the circles in the top
right panel.

\subsection{The remarkably flat disk of NGC 3628}
\label{optical3628}

As noted above, NGC 3628 has an asymmetic distribution of $8\;\mu$m cores around the
midplane with a thicker disk in the west because of its interaction with other
galaxies in the Leo triplet.  The disk of cores is still very thin and straight,
however, compared to the extended dust. Figure
\ref{N3628ch4divg3_APOD_5-15-08_overlaynice_northup_double} shows a deep color image
of NGC 3628 \citep[Astronomy Picture of the Day from May 15, 2001, credit: Keith
Quattrocchi][]{} with the unsharp mask image from Figure 2 superposed to scale. The
extreme flatness of the $8\;\mu$m disk, even in an interacting galaxy, presumably
results from a high degree of stability in the perpendicular direction, as suggested
for spiral galaxies in general by \cite{saha06}. Very few of the $8\;\mu$m cores are
at the positions of optical features, although some at high latitude in Figure
\ref{n3628_usm_dual_v2} may be seen in the dust loop that goes above the disk
between the center and the western part.

\begin{figure}
\epsscale{1.}
\includegraphics[width=6.in]{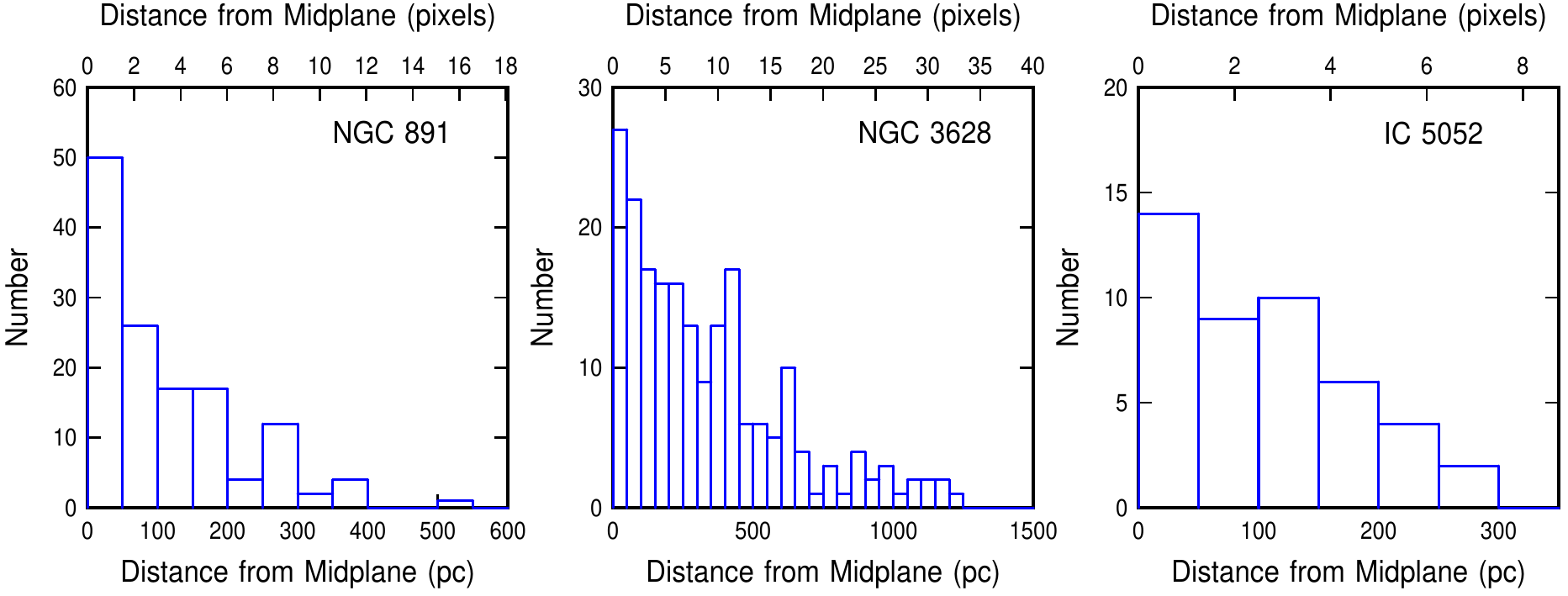}
\caption{Distribution functions of $8\;\mu$m core distance perpendicular to the midplane,
in units of pc on the lower scale and pixels on the upper scale. The bin size is 50 pc. The dispersion
in these distances is the disk half-thickness; from left to right it is: 105 pc, 440 pc, and 74 pc.
Half the disk of NGC 3628 is puffed up because of an interaction, making the average thickness
large. }
\label{efremov_edgeon_rms_pc}
\end{figure}

\begin{figure}
\epsscale{1.}
\includegraphics[width=6.in]{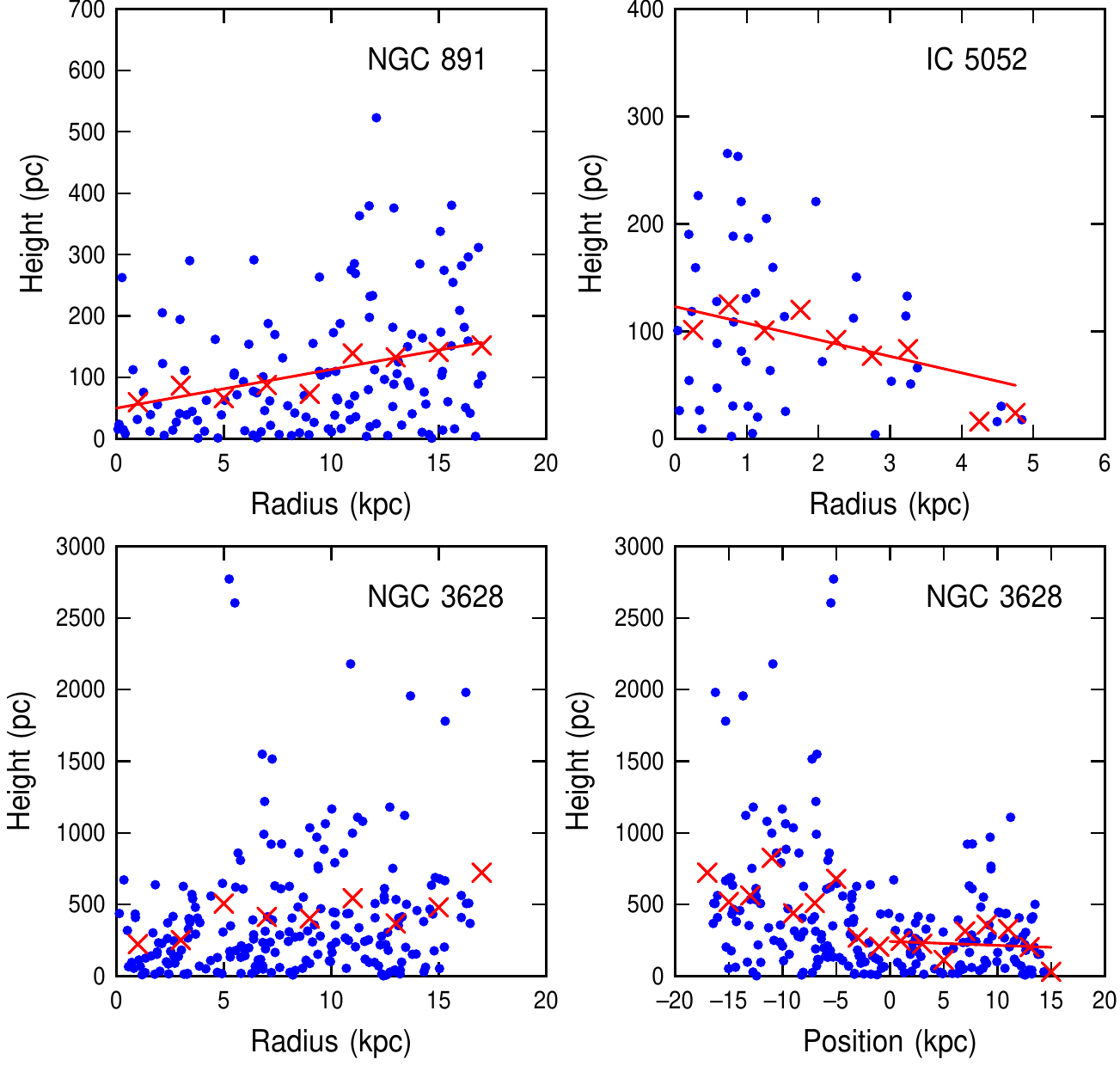}
\caption{Distribution of distance perpendicular to the midplane (``height'')
versus galactocentric radius for each galaxy and versus position along the major
axis in the case of NGC 3628 (lower right panel) to distinguish the distorted portion
of the disk (negative position, which is to the west). Blue points represent each clump
and red crosses are the average values in bins around them.  Linear fits to the height
versus radius are shown; for NGC 3628, the fit avoids the thicker region in the west that
was probably affected by the interaction. }
\label{efremov_edgeon_xy2}
\end{figure}

\begin{figure}
\epsscale{1.}
\includegraphics[width=6.in]{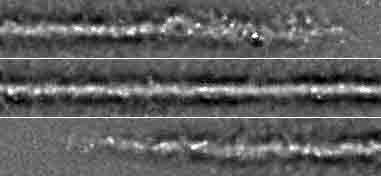}
\caption{Unsharp mask image NGC 891 at $8\;\mu$m
is divided into three segments to highlight details of the
perpendicular displacement of $8\;\mu$m cores from the midplane.
The north-east end of the disk
is in the bottom panel and the south-west end in the top panel (the disk was rotated
counter-clockwise by $67^\circ$ to make it horizontal). (Figure degraded for arXiv storage.)
}
\label{n891_separated}
\end{figure}

\begin{figure}
\epsscale{1.}
\includegraphics[width=6.in]{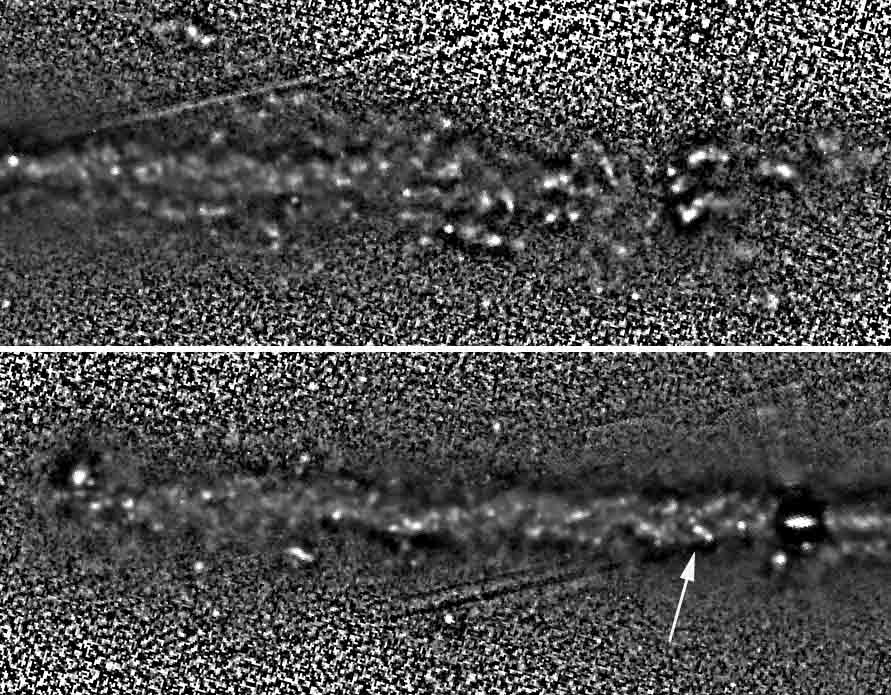}
\caption{Two segments of the unsharp mask image of NGC 3628 at $8\;\mu$m
rotated clockwise by $15^\circ$ to highlight the perpendicular
structures. The eastern half is in the bottom panel.
To the left of the nuclear disk, in the lower panel, there is a pattern of 3 arcs
symmetrically placed around the midplane (the inner one is highlighted by
an arrow); these arcs could be spiral arms. (Figure degraded for arXiv storage.)}
\label{n3628_separated}
\end{figure}

\begin{figure}
\epsscale{1.}
\includegraphics[width=6.in]{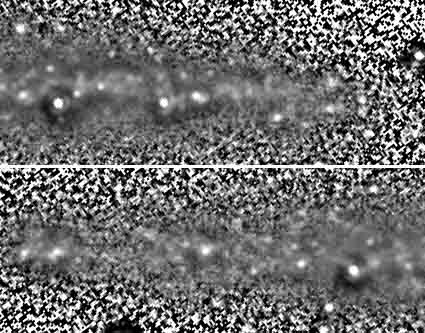}
\caption{Two segments of the unsharp mask image of IC 5052 at $8\;\mu$m
rotated clockwise by $54^\circ$ to highlight the perpendicular
structures. The eastern half is in the bottom panel. The upper panel has two linear streaks of 3 or 4 cores and the lower
panel has one fainter, and tilted in the other direction. (Figure degraded for arXiv storage.)}
\label{ic5052_separated}
\end{figure}

\begin{figure}
\epsscale{1.}
\includegraphics[width=6.in]{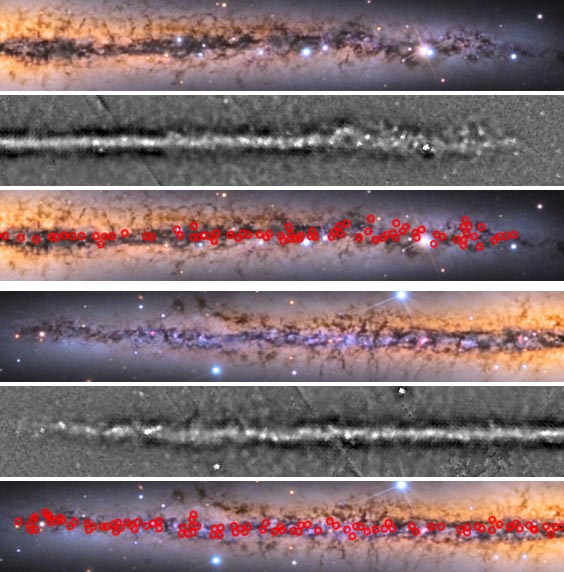}
\caption{Two sections of NGC 891 showing a comparison between an optical
image of NGC 891 from Astronomy Picture of the Day, January 12, 2017,
(credit: Adam Block), the unsharp mask image at $8\;\mu$m, and the positions of the
catalogued $8\;\mu$m cores as circles superposed on the optical image.
The north-east end of the disk
is in the bottom panel and the south-west end in the top panel (the disk was rotated
counter-clockwise by $67^\circ$ to make it horizontal). Only a few of the circles
contain optical emission, which means that most of the $8\;\mu$m cores are invisible
optically. (Figure degraded for arXiv storage.)}
\label{N891APOD_ch4divg3_triplet_thin_double_v2}
\end{figure}

\begin{figure}
\epsscale{1.}
\includegraphics[width=6.in]{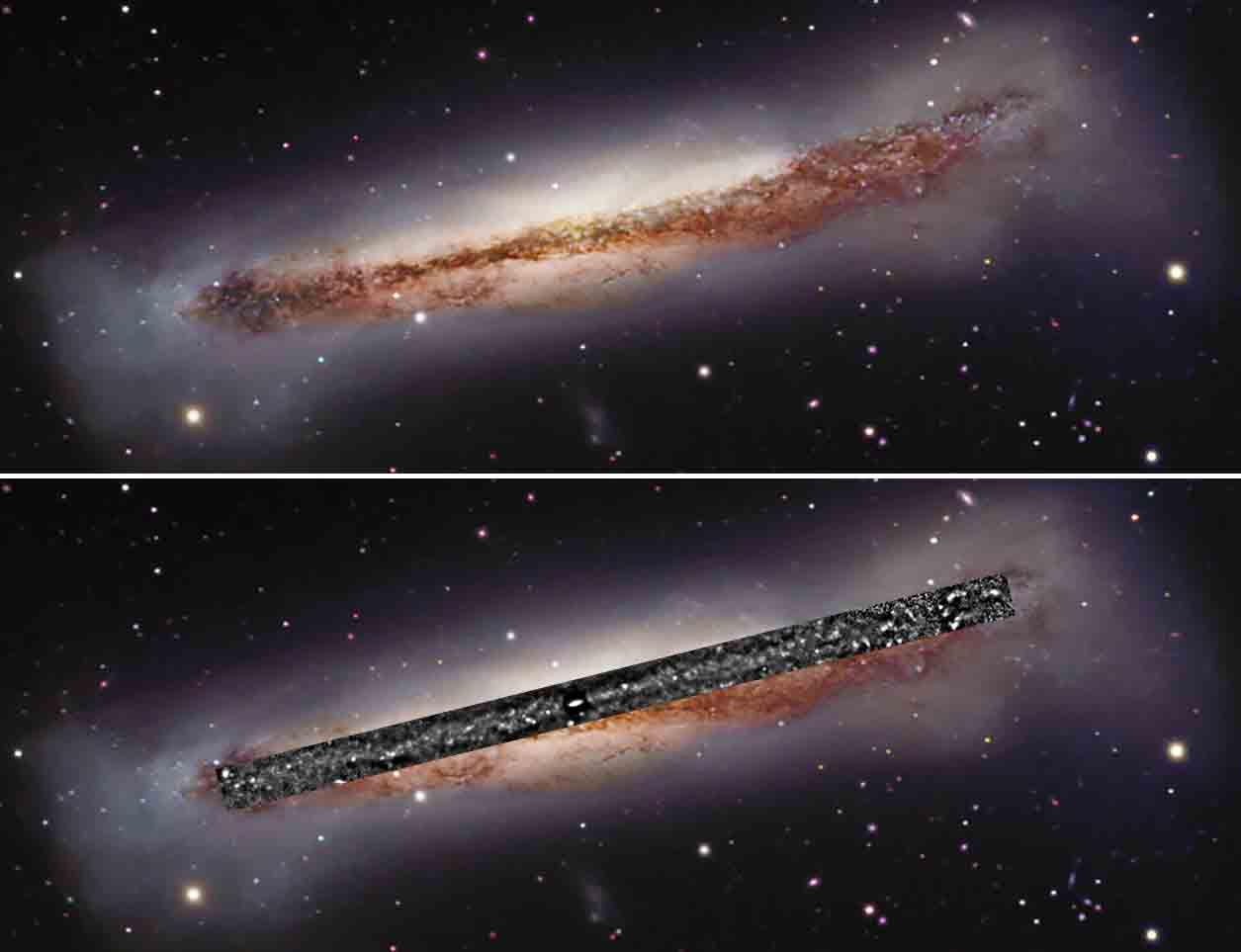}
\caption{(top) Color image of NGC 3628 from Astronomy Picture of the Day, May 15, 2001
(credit: Keith Quattrocchi) and (bottom) the same image
with the thin disk part of the $8\;\mu$m unsharp mask superposed at the correct position.
In spite of the irregularity
and high displacement of the optical emission due to this galaxy's interaction in the
Leo Triplet, the disk of $8\;\mu$m cores is remarkably flat and thin, suggesting great stability
even though it thickens at the ends. (Figure degraded for arXiv storage.)
}
\label{N3628ch4divg3_APOD_5-15-08_overlaynice_northup_double}
\end{figure}

The optical image thickens or flares at each end of the disk but the core
distribution flares primarily in the west (Fig. \ref{efremov_edgeon_rms_pc}),
although it is thicker everywhere than the core distributions in the other galaxies.
The thick optical distribution in IC 5052 could be dominated by stars at large
radii, which would be perturbed more than the denser inner disk.

\subsection{IC 5052}
\label{color5052}

\begin{figure}
\epsscale{1.}
\includegraphics[width=6.in]{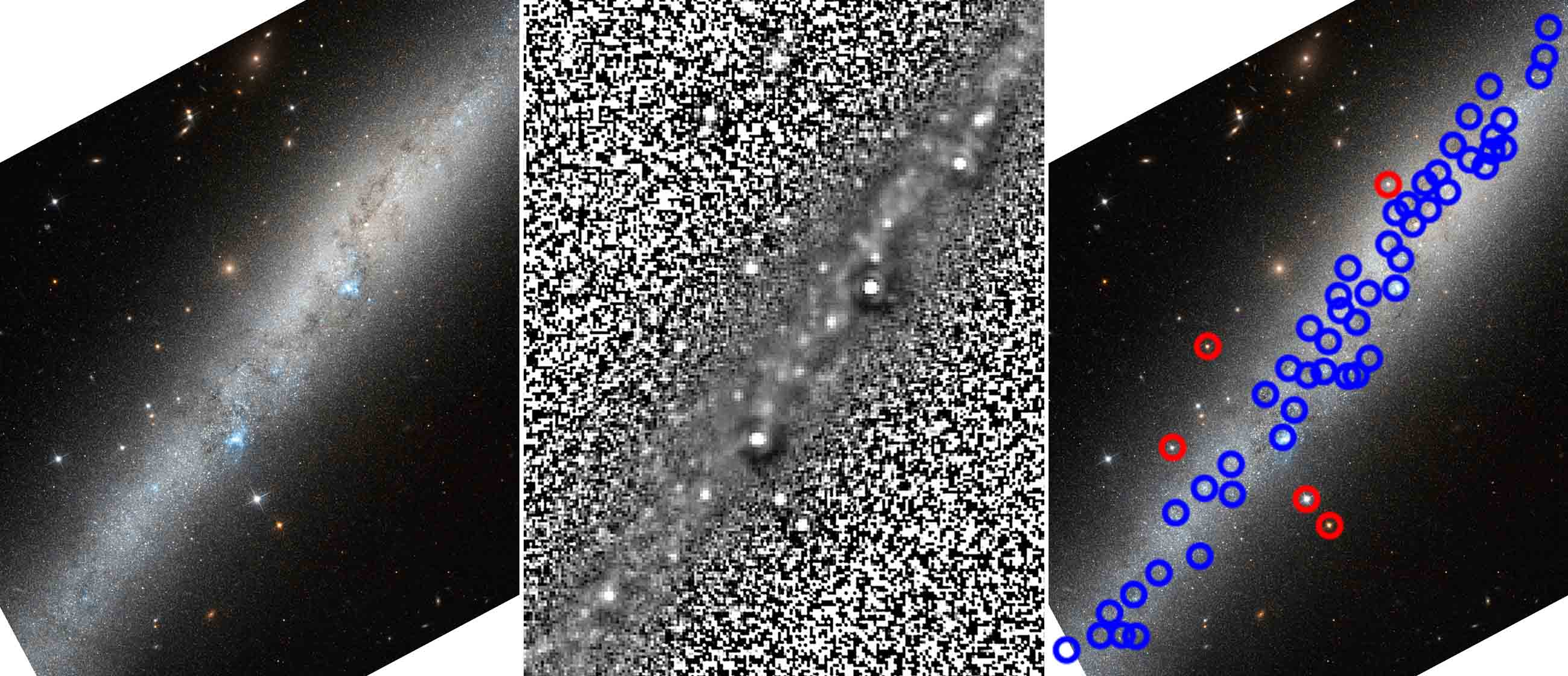}
\caption{(left) Hubble Space Telescope image of the inner part of IC 5052
(credit:  ESA/Hubble and NASA, S.
Meunier);
(middle) unsharp mask image at
$8\;\mu$m as in Figure \ref{n3628_usm_dual_v2}; (right) superposition of the
$8\;\mu$m cores as in Figure \ref{n3628_usm_dual_v2}, but now drawn as empty blue circles
to show the associated optical structure from the Hubble image, along with red
circles to show stars that were useful for image alignment. The two bright blue
regions symmetric around the center and slightly below the midplane in the optical
image on the left correspond to the two brightest $8\;\mu$m cores in the unsharp mask
image. These regions are highlighted by red circles in Fig. \ref{efremov_edgeon_colors}
and they correspond to the two brightest cores in the distribution function of
Fig. \ref{efremov_edgeon_mags}. (Figure degraded for arXiv storage.)}
\label{Hubble_I5052_triplet}
\end{figure}

Figure \ref{Hubble_I5052_triplet} compares an optical image of IC 5052 from the
Hubble Space
Telescope\footnote{https://www.nasa.gov/mission\_pages/hubble/science/ic5052.html}
with the $8\;\mu$m image studied here. This is the closest of the three galaxies and
shows the most detailed features from star formation. The left-hand panel is the HST
image, the middle panel is the unsharp mask image from Figure
\ref{n3628_usm_dual_v2}, and the right hand image has the positions of the
catalogued $8\;\mu$m cores outlined as blue circles, with stars for alignment
outlined are red circles. The two bright blue regions symmetric around the center
and slightly below the midline of the galaxy are also $8\;\mu$m cores, as indicated
by the presence of blue optical emission inside the blue circles in the right-hand
panel. A few other $8\;\mu$m core regions contain optical emission also, but most do
not. Some cores correspond to dark clouds in the optical image, such as the two
symmetrically placed between the bright blue regions and several at high latitude to
the northwest. Most of the cores show no clear indication of any optical feature.

IRAC colors of the two bright blue regions in the Hubble image are indicated by red
circles in Figure \ref{efremov_edgeon_colors}. They have redder than average
$[3.6]-[4.5]$ colors suggesting high extinction and/or hotter dust in a deeply
embedded, optically-invisible core. They are also the brightest $8\mu$m cores in the
unsharp mask image (Fig. \ref{Hubble_I5052_triplet}). Their apparent magnitudes are
$10.59\pm0.09$ and $11.20\pm0.11$. They are the two cores that stand out at the
bright magnitude side of the distribution in Figure \ref{efremov_edgeon_mags}. The
cores themselves are not evident as dark clouds in the optical images.

\section{Star formation in $8\;\mu$m cores}

The IRAC luminosities of the cores were converted to stellar masses using the
procedure in Papers 1 and 2. We first multiply the sum of the luminosities of the
four IRAC bands, $L_{\rm IRAC}$, by 10 to get an estimate of the total infrared
luminosity, based on starburst SEDs in \cite{xu01}. Then the IR luminosity is
converted to young stellar mass using the bolometric magnitude of a young stellar
population in \cite{bruzual03}, which is $-2.2362$ for one solar mass at solar
metallicity, an age less than or equal to 1 Myr, and a Salpeter initial mass
function (as used to determine the star formation rates in Table 1). Combined with
the bolometric magnitude of the Sun, 4.74 mag, this gives the total luminosity per
unit solar mass of young stars in erg s$^{-1}$. After solving for mass, we have the
stellar masses associated with each core,
\begin{equation}
M_{\rm stars}(M_\odot) \sim {{10L_{\rm IRAC}({\rm erg\;s}^{-1})}
\over{3.828\times10^{33}\times10^{0.4\times[4.74+2.2362]}}}.
\label{mass}
\end{equation}

The IRAC luminosities were increased over the observed values to account for
extinction in the infrared, assuming an average of $A_{\rm V}=20$ mag for each
source, foreground and local combined (see above). To do this we use the conversions
in \cite{chapman09} and \cite{bessell88}, which give: $A_{3.6}/A_{\rm V}=0.06976$,
$A_{4.5}/A_{\rm V}=0.05777$, $A_{5.8}/A_{\rm V}=0.05014$, and $A_{8.0}/A_{\rm
V}=0.04905$. The extinctions in the IRAC bands are then around 1 magnitude. They
have the effect of increasing the stellar masses and effective ages of the regions
by about a factor of 2.5 compared to the case with no infrared extinction. Analogous
extinctions were not considered for stellar masses in Papers 1 and 2 because the
total extinctions were less by about half. Including them would have increased the
masses and ages by $\sim60$\%.

For the extinction-corrected IRAC luminosities in equation \ref{mass}, the total
stellar masses associated with the catalogued $8\;\mu$m cores in each galaxy range
from $10^5\;M_\odot$ to $4\times10^6\;M_\odot$, as listed in Table \ref{masses}.
Dividing these by the total galactic star formation rates (Table \ref{masses}) under
the assumption that they represent all the star-forming regions gives average
lifetimes from 0.8 Myr to 2 Myr.

The star formation rates for NGC 891 and NC 3628 are from \cite{shinn15} using IRAS
far-infrared observations with the calibration in \cite{kennicutt98} and corrected
for the distances given here. For NGC 891, \cite{popescu11} get a similar SFR using
a radiative transfer fit to the SED; their rate converts to $2.65\;M_\odot$
yr$^{-1}$ for our distance. Another radiative transfer model for NGC 891 in
\cite{seon14} gets a rate of 3-4 $M_\odot$ yr$^{-1}$ using Starburst 99 and a
distance of 9.5 Mpc. The rate for IC 5052 uses the extinction-corrected H$\alpha$
flux in \cite{kaisin07} with the calibration in \cite{kennicutt98} and the distance
given here. All star formation rates assume a Salpeter IMF.

The average stellar masses of individual cores range from $2\times10^3\;M_\odot$ to
$3\times10^4\;M_\odot$ for the three galaxies. These masses do not account for
completeness limits, so a better measure is the mass at the peak of the luminosity
function shown in Figure \ref{efremov_edgeon_mags}. Again considering ages of 1 Myr,
a Salpeter IMF, and the summed extinction-corrected IRAC luminosities as a proxy for
the total luminosities, the average masses within one magnitude range around the
peaks in the luminosity functions range between $10^3\;M_\odot$ and
$1.7\times10^4\;M_\odot$ (Table \ref{masses}). Lower mass regions are missing
because of faintness. Correcting for these missing cores would approximately double
the total stellar mass and lifetime (Paper 2).

\section{Conclusions}
\label{conc}

Unsharp masks of $8\;\mu$m IRAC images show several hundred emission sites in three
highly inclined galaxies.  The cores have IRAC colors and magnitudes typical of
star-forming regions with an average intrinsic extinction of $\sim8.8$ mag, based on
the results for face-on galaxies in Paper 2, and a corresponding gas mass surface
density of $\Sigma_{\rm gas}\sim95\;M_\odot$ pc$^{-2}$. Foreground extinction
averages $\sim11.2$ mag through the edge-on disks. The average young stellar mass in
each region is $\sim10^4\;M_\odot$, depending on galaxy distance. If the local star
formation efficiency is 5\%, then this mass and extinction imply an average cloud
core diameter of $\sim18$ pc. The ratio of the total core mass to the star formation
rate in each galaxy gives a lifetime for the highly extincted region of 1 to 2 Myr.

The identification of young star-forming regions allows an assessment of where stars
form relative to galactic-scale processes. In Papers 1 and 2, we found that most
young regions lie along filaments that span several kpc in length and run parallel
to spiral arms and spurs. These filaments are presumably shock fronts connected with
gas flows in spiral density waves, and the cores are gravitational condensations in
the shock fronts.  Other cores are in kpc-scale rings that could be made by young
stellar feedback. The present paper reveals the vertical distribution of the young
regions. The disks studied here have half-thicknesses of 105 pc for NGC 891 and 74
pc for IC 5052, which are typical of molecular layers in normal galaxies. The
half-thickness is 438 pc for NGC 3628, but that includes a broad region affected by
a galaxy interaction.  There is also a loop of several cores above the disk in NGC
891 and possibly coherent loops of cores in the thick part of NGC 3628.

Some of the $8\;\mu$m cores show coherent structures suggestive of kpc-scale
corrugations, shingles, or spiral arms. For example, NGC 3628, which is slightly
less inclined than the other galaxies, has arcs composed of several cores each that
could be spiral arms viewed tangentially. The spacing between the cores in these
arcs is about the same as the spacing between $8\;\mu$m cores in the spiral arms of
M100, studied in Paper 1. Optical images of NGC 3628 show no evidence for these
spirals because of extinction. Most of the cores have no trace in the optical images
anyway, even in face-on galaxies (Paper 1).  This invisibility highlights the
importance of the $8\;\mu$m cores as tracers of an early stage in star formation.

Disk thicknesses for the $8\;\mu$m cores are either somewhat constant or increase
slightly with radius. Because this thickness is directly related to the youngest
stages of star formation, its observation may be useful to help us understand the
sources of interstellar turbulence before molecular clouds form. The near-constant
thickness implies that the effective vertical velocity dispersion, $\sigma$,
decreases with galactocentric distance slightly slower than the square root of the
total surface density, $\Sigma$, considering that the equation for height of an
isothermal layer is $H=\sigma^2/(2\pi G\Sigma)$. That translates to a factor of 1.6
in $\sigma$ for each scale length in the disk. Radial decreases in $\sigma$ have
been observed in face-on galaxies \citep[e.g.,][]{dickey90}. A square root
correlation like this was seen directly in face-on Ultra Luminous Infrared Galaxies
\citep{wilson19}.

The observation of a slowly varying thickness is also useful to convert the disk
surface densities that are usually measured in galaxies into volume densities, which
are important to get timescales for gravitationally-driven processes.  As is well
known, the $\sim1.5$ slope of the Kennicutt-Schmidt relation for total gas is
trivially explained if the thickness of the star-forming part of the disk varies
more slowly than the surface density, as is the case here. Then the midplane density
that determines the timescale for large-scale gravitation collapse is nearly
proportional to the surface density, and the square root of this midplane density,
which gives the collapse rate and presumably also the star formation rate per unit
gas mass, is proportional to the square root of the surface density
\citep{madore77,elmegreen18b}. The total star formation rate per unit area is
proportional to the gas mass per unit area times the collapse rate.

This work is based in part on observations made with the {\it Spitzer Space
Telescope}, which is operated by the Jet Propulsion Laboratory, California Institute
of Technology under a contract with NASA. We are grateful to Dr. Tom Megeath for
help with the conversion from IRAC color to extinction, and to the referee for
useful comments.

\newpage
\begin{deluxetable}{lccccccc}
\tabletypesize{\scriptsize} \tablecolumns{8} \tablewidth{0pt}
\tablecaption{Galaxies} \tablehead{
\colhead{Name}&
\colhead{Dist.\tablenotemark{a}}&
\colhead{Dist.}&
\colhead{Incl.\tablenotemark{b}}&
\colhead{Hubble type\tablenotemark{c}}&
\colhead{$R_{25}$\tablenotemark{d}} &
\colhead{$R_{25}$} &
\colhead{pc per}
\\
\colhead{}&
\colhead{Mpc}&
\colhead{Modulus}&
\colhead{deg}&
\colhead{}&
\colhead{arcmin}&
\colhead{kpc}&
\colhead{arcsec}
}
\startdata
NGC 891	    &	9.12	&	29.8	&   89.7   & 	SA(s)b III	    &		6.74	&	17.9	&	44.2 \\
NGC 3630	&	10.3    &	30.7	&	79.3   &    Sb pec III   	&		7.40	&	22.2	&	50.5 \\
IC 5052 	&	5.5 	&	28.7	&	90   &      SBd VII     	&	    2.94	&	4.70	&	26.7 \\
\enddata
\tablenotetext{a}{Distances from \cite{tully13}, which are consistent with a Hubble constant of $74.4\pm3.0$}
\tablenotetext{b}{Inclination for NGC 891 is from \cite{xilouris98}, and for NGC 3628 and IC 5052 they are from HyperLeda, http://leda.univ-lyon1.fr/}
\tablenotetext{c}{Hubble types from \cite{devauc91}}
\tablenotetext{d}{$R_{\rm 25}$ from \cite{devauc91}}
\label{galaxylist}
\end{deluxetable}

\newpage
\begin{deluxetable}{lcccc}
\tabletypesize{\scriptsize} \tablecolumns{4} \tablewidth{0pt}
\tablecaption{Measurement Data} \tablehead{
\colhead{Band}&
\colhead{Ang. Resol.\tablenotemark{a}}&
\colhead{Zero Pt.\tablenotemark{b}}&
\colhead{Aper. Corr.\tablenotemark{c}}\\
\colhead{$\mu$m}&
\colhead{Arcsec FWHM}&
\colhead{Counts}&\\
}
\startdata
3.6	    &	1.1     &   277.5	&	1.215	\\
4.5 	&	1.3     &   179.5   &	1.233	\\
5.8 	&	1.7     &   116.6 	&	1.366	\\
8.0     &   2.4     &    63.1   &   1.568   \\
\enddata
\tablenotetext{a}{Angular resolution from \cite{chambers09}; see also \cite{fazio04}}
\tablenotetext{b}{Zero magnitude fluxes are from the Fits file headers and also Table 4.1 in the IRAC Instrument Handbook, https://irsa.ipac.caltech.edu/data/SPITZER/docs/irac/iracinstrumenthandbook/17/}
\tablenotetext{c}{Aperture corrections are for the smallest point sources in
Table 4.7 of the IRAC Instrument Handbook, https://irsa.ipac.caltech.edu/data/SPITZER/docs/irac/iracinstrumenthandbook/27/}
\label{photometry}
\end{deluxetable}

\newpage
\begin{deluxetable}{lcccccccc}
\tabletypesize{\scriptsize} \tablecolumns{8} \tablewidth{0pt}
\tablecaption{Average Colors and their Dispersions}
\tablehead{
\colhead{Galaxy}&
\colhead{N\tablenotemark{a}}&
\colhead{Mask}&
\colhead{$<[3.6]-[4.5]>$}&
\colhead{$\Delta([3.6]-[4.5])$\tablenotemark{b}}&
\colhead{$<[4.5]-[5.8]>$}&
\colhead{$\Delta([4.5]-[5.8]$)}&
\colhead{$<[5.8]-[8.0]>$}&
\colhead{$\Delta([5.8]-[8.0]$)}\\
\colhead{}&
\colhead{}&
\colhead{pc}&
\colhead{mag}&
\colhead{mag}&
\colhead{mag}&
\colhead{mag}&
\colhead{mag}&
\colhead{mag}
}
\startdata
NGC 891	  & 133  &	199 & $0.42\pm0.05$ & 0.52 & $2.14\pm0.06$  & 0.64 & $1.75\pm0.03$ & 0.33 \\
NGC 3628  & 211  &  225 & $0.31\pm0.03$ & 0.39 & $2.07\pm0.05$  & 0.68 & $1.89\pm0.02$ & 0.30 \\
IC 5052   & 45   &  120 & $0.32\pm0.06$ & 0.42 & $1.61\pm0.12$  & 0.80 & $1.83\pm0.06$ & 0.40 \\
All       & 389  &      & $0.35\pm0.02$ & 0.45 & $2.04\pm0.04$  & 0.70 & $1.83\pm0.02$ & 0.33 \\
\enddata
\tablenotetext{a}{The number of cores includes only those with all three
measured colors. Cores with one or more undetected passbands are not here, although
they are in the core tabulations that follow. These numbers are 173, 267 and 60 in the order above.}
\tablenotetext{b}{The color $\Delta$ is the dispersion in the measured color distribution.}
\label{colors}
\end{deluxetable}

\newpage
\begin{deluxetable}{lcccccc}
\tabletypesize{\scriptsize} \tablecolumns{8} \tablewidth{0pt}
\tablecaption{Equivalent Stellar Masses and Ages}
\tablehead{
\colhead{Galaxy}&
\colhead{SFR\tablenotemark{a}}&
\colhead{Total Measured}&
\colhead{Eff. Age\tablenotemark{b}}&
\colhead{Average Clump}&
\colhead{Peak Absolute}&
\colhead{Ave. Mass }
\\
\colhead{}&
\colhead{$M_\odot$ yr$^{-1}$}&
\colhead{Clump Mass, $M_\odot$}&
\colhead{Myr}&
\colhead{Mass, $M_\odot$}&
\colhead{Mag Range}&
\colhead{Around Peak\tablenotemark{c} ($M_\odot$)}
}
\startdata
NGC 891	  & 2     & $4.45\times10^6$  & 2.3 & $3.3\times10^4$ & $-17$ to $-18$ & $1.7\pm0.9\times10^4$ \\
NGC 3628  & 4.2   & $3.33\times10^6$  & 0.79 & $1.6\times10^4$ & $-16$ to $-17$ & $6.4\pm4.9\times10^3$ \\
IC 5052   & 0.056 & $9.49\times10^4$  & 1.7 & $2.1\times10^3$ & $-14$ to $-15$ & $1.0\pm0.3\times10^3$   \\
\enddata
\tablenotetext{a}{Star formation rates for NGC 891 and NC 3628 are from
\cite{shinn15} using IRAS far infrared observations. IC 5052 uses the
extinction-corrected H$\alpha$ flux in \cite{kaisin07}. All star formation rates
assume a Salpeter IMF and were converted to the distances assumed here.}
\tablenotetext{b}{The effective age is the ratio of the total stellar mass
associated with the $8\;\mu$m cores to the star
formation rate in the galaxy.} \tablenotetext{c}{The average mass around the peak is the average
within the absolute magnitude interval at $8\;\mu$m given in the previous column.}
\label{masses}
\end{deluxetable}

\newpage
\begin{deluxetable}{llccccccc}
\tabletypesize{\scriptsize} \tablecolumns{9} \tablewidth{0pt}
\tablecaption{$8\;\mu$m Cores in NGC 891\tablenotemark{a}} \tablehead{
\colhead{RA\tablenotemark{b}}& \colhead{DEC}& \colhead{[3.6]\tablenotemark{c}}&
\colhead{[4.5]}& \colhead{[5.8]} & \colhead{[8.0]} & \colhead{$[3.6]-[4.5]$} &
\colhead{$[4.5]-[5.8]$} &
\colhead{$[5.8]-[8.0]$} \\
} \startdata
 2:22:43.90& 42:25:42.9&$  19.18\pm   1.82$&$        -        $&$        -        $&$  16.46\pm   1.28$&$  -  $&$  -  $&$  -  $\\
 2:22:43.85& 42:25:36.3&$  19.25\pm   1.88$&$  18.69\pm   1.80$&$        -        $&$  17.80\pm   2.97$&   0.54&$  -  $&$  -  $\\
 2:22:44.06& 42:25:34.5&$  18.80\pm   1.51$&$  18.94\pm   2.03$&$  18.13\pm   1.93$&$  16.23\pm   1.20$&  -0.16&   0.70&   1.75\\
 2:22:43.94& 42:25:31.9&$  18.97\pm   1.70$&$  18.95\pm   2.06$&$  17.78\pm   1.73$&$  15.27\pm   0.73$&   0.01&   1.05&   2.36\\
 2:22:43.52& 42:25:35.5&$        -        $&$        -        $&$  17.85\pm   1.60$&$  16.17\pm   1.12$&$  -  $&$  -  $&   1.54\\
... &&&&&&&&\\
\enddata
\tablenotetext{a}{This table lists 173 cores visible on an unsharp mask image of the
galaxy at $8\;\mu$m. As suggested by Figure \ref{efremov_edgeon_mags}, the
completion limit for apparent magnitude is around 13.} \tablenotetext{b}{In order of
RA} \tablenotetext{c}{Magnitudes in the Vega scale} \label{listofclumps891}
\end{deluxetable}
\newpage

\begin{deluxetable}{llccccccc}
\tabletypesize{\scriptsize} \tablecolumns{9} \tablewidth{0pt}
\tablecaption{$8\;\mu$m Cores in NGC 3628} \tablehead{ \colhead{RA}& \colhead{DEC}&
\colhead{[3.6]}& \colhead{[4.5]}& \colhead{[5.8]} & \colhead{[8.0]} &
\colhead{$[3.6]-[4.5]$} & \colhead{$[4.5]-[5.8]$} &
\colhead{$[5.8]-[8.0]$} \\
} \startdata
11:20:36.09& 13:34:13.5&$  19.43\pm   2.01$&$  18.36\pm   1.54$&$  16.99\pm   1.10$&$  15.66\pm   1.02$&   1.04&   1.26&   1.18\\
11:20:35.15& 13:34:26.4&$  15.40\pm   0.32$&$  14.96\pm   0.33$&$  12.94\pm   0.18$&$  11.19\pm   0.13$&   0.43&   1.90&   1.61\\
11:20:34.85& 13:34:25.7&$  16.64\pm   0.56$&$  16.36\pm   0.62$&$  14.28\pm   0.36$&$  12.33\pm   0.26$&   0.26&   1.97&   1.79\\
11:20:35.64& 13:34:17.6&$  18.71\pm   1.45$&$  18.65\pm   1.77$&$  16.08\pm   0.73$&$  13.96\pm   0.40$&   0.04&   2.47&   1.97\\
11:20:35.41& 13:34:18.4&$  18.62\pm   1.42$&$  18.45\pm   1.66$&$  15.87\pm   0.71$&$  13.72\pm   0.39$&   0.15&   2.47&   2.00\\
...&&&&&&&&\\
\enddata
\tablenotetext{a}{This table lists 267 cores visible on an unsharp mask image of the
galaxy at $8\;\mu$m. As suggested by Figure \ref{efremov_edgeon_mags}, the
completion limit for apparent magnitude is around 15.} \tablenotetext{b}{In order of
RA.} \tablenotetext{c}{Magnitudes in the Vega scale.} \label{listofclumps3628}
\end{deluxetable}
\newpage

\begin{deluxetable}{llccccccc}
\tabletypesize{\scriptsize} \tablecolumns{9} \tablewidth{0pt}
\tablecaption{$8\;\mu$m Cores in IC 5052} \tablehead{ \colhead{RA}& \colhead{DEC}&
\colhead{[3.6]}& \colhead{[4.5]}& \colhead{[5.8]} & \colhead{[8.0]} &
\colhead{$[3.6]-[4.5]$} & \colhead{$[4.5]-[5.8]$} &
\colhead{$[5.8]-[8.0]$} \\
} \startdata
20:52:22.55&-69:14: 6.8&$  19.28\pm   1.91$&$  18.20\pm   1.44$&$  16.42\pm   0.84$&$  14.49\pm   0.50$&   1.06&   1.67&   1.79\\
20:52:21.24&-69:13:58.4&$  17.62\pm   0.89$&$  17.25\pm   0.93$&$  15.49\pm   0.55$&$  13.77\pm   0.38$&   0.35&   1.66&   1.57\\
20:52:21.28&-69:13:55.7&$  17.87\pm   0.99$&$  17.50\pm   1.04$&$  15.40\pm   0.52$&$  13.51\pm   0.32$&   0.35&   1.99&   1.75\\
20:52:18.49&-69:13:25.0&$        -        $&$        -        $&$  17.62\pm   1.44$&$  15.57\pm   0.83$&$  -  $&$  -  $&   1.90\\
20:52:16.83&-69:13:21.3&$  19.70\pm   2.32$&$  19.58\pm   2.74$&$  18.10\pm   1.82$&$  15.63\pm   0.87$&   0.10&   1.37&   2.32\\
...&&&&&&&&\\
\enddata
\tablenotetext{a}{This table lists 60 cores visible on an unsharp mask image of the
galaxy at $8\;\mu$m. As suggested by Figure \ref{efremov_edgeon_mags}, the
completion limit for apparent magnitude is around 15.} \tablenotetext{b}{In order of
RA} \tablenotetext{c}{Magnitudes in the Vega scale} \label{listofclumps5052}
\end{deluxetable}

\end{document}